\documentclass[11pt]{article}
\usepackage[utf8]{inputenc}

\usepackage{jheppub}

\usepackage{soul}
\usepackage[figtopcap]{subfigure}
\usepackage{amsmath,amssymb,dsfont}
\usepackage{hyperref}
\usepackage[T1]{fontenc} 
\usepackage{txfonts}
\usepackage{newlfont}
\usepackage{times}

\newcommand\Ei{\,\textmd{Ei}}

\title{   $f(T)$ cosmology: From Pseudo-Bang  to Pseudo-Rip}
\author[a]{W. El Hanafy}
\author[b,c,d]{and Emmanuel N. Saridakis}

\affiliation[a]{Centre for Theoretical Physics, The British University in
Egypt, P.O. Box 43, El Sherouk City, Cairo 11837, Egypt}
 \affiliation[b]{National Observatory of Athens, Lofos Nymfon, 11852 Athens,
Greece}

  \affiliation[c]{CAS Key Laboratory for Researches in Galaxies and Cosmology,
Department of Astronomy, University of Science and Technology of China, Hefei,
Anhui 230026, P.R. China}
 \affiliation[d]{School of Astronomy, School of Physical Sciences,
University of Science and Technology of China, Hefei 230026, P.R. China}

\emailAdd{waleed.elhanafy@bue.edu.eg}
\emailAdd{msaridak@noa.gr}

\abstract{We   investigate  the complete universe evolution in the framework of
$f(T)$ cosmology. We first study the  requirements at the
kinematic level and we introduce a simple scale factor with the necessary
features. Performing a detailed analysis of the   phase
portrait  we show that   the universe   begins   in the infinite past  from a
phase where the
scale factor goes to zero but  the Hubble parameter goes to a constant, and its
derivative to zero. Since these features resemble those of the Pseudo-Rip
fate   but in a reverted way, we call this initial phase
as  Pseudo-Bang. Then  the universe evolves in a first inflationary phase, a
cosmological turnaround and a bounce, after which we have a
second inflationary regime with a successful exit. Subsequently we obtain
the standard thermal history and the sequence of radiation, matter and
late-time acceleration epochs, showing that  the  universe will result in
an everlasting Pseudo-Rip phase. Finally, taking  advantage of the fact that
the field equations of $f(T)$ gravity are of second order, and therefore the
corresponding autonomous dynamical  system is  one dimensional,
  we incorporate the aforementioned kinematic
features and we reconstruct  the specific $f(T)$
form that can dynamically generate the  Pseudo-Bang cosmological scenario.
 Lastly, we examine  the evolution of the primordial fluctuations
showing that they  are initially  sub-horizon, and we show that
the total fluid does not exhibit any singular behaviour at the
phantom  crossing points, while the   torsional fluid experiences
them as Type II singular phases.}
\begin{document}
\maketitle
\flushbottom

\section{Introduction}\label{Sec:1}

  According to accumulating observational evidence from different and various
probes, the universe passed through
two phases of accelerated expansion, one at very early and one at late
cosmological times. In order to explain these phases one needs to proceed to a
modification of the standard lore of cosmology.
A first direction he could follow is to introduce new particles/fields while
still remaining in the framework of general relativity, such as the
inflaton field
\cite{Olive:1989nu,Bartolo:2004if}  and
the dark energy sector
\cite{Copeland:2006wr,Cai:2009zp}.
A second direction is to construct  gravitational modifications, which deviate
from general relativity at particular scales,   thus offering the
 extra degrees of freedom needed to describe the universe evolution
\cite{Capozziello:2011et,CANTATA:2021ktz}.

In order to build a modified gravity theory one modifies a specific feature of
general relativity.   Altering
the   Einstein-Hilbert Lagrangian leads to
  $f(R)$ gravity
\cite{DeFelice:2010aj,Nojiri:2010wj}, Modified Gauss-Bonnet theory \cite{Nojiri:2005jg}, Lovelock gravity
\cite{Lovelock:1971yv}, etc, while allowing for a scalar field
 coupled with curvature invariants    gives rise to
Horndeski \cite{Horndeski:1974wa} and generalized Galileon
theories \cite{Nicolis:2008in,Deffayet:2009wt}. Additionally, changing  the
spacetime
dimensionality  leads  to the braneworld theories
\cite{Brax:2004xh}. An interesting alternative is to start
 from the equivalent, teleparallel formulation of gravity
\cite{ein28,ein28a,ein28b,Hayashi79,Pereira.book,Maluf:2013gaa} and construct
modifications
using combinations of torsional
invariants, such as in    $f(T)$ gravity
\cite{Cai:2015emx,Bengochea:2008gz,Linder:2010py},   in
$f(T,T_G)$ gravity
\cite{Kofinas:2014owa,Kofinas:2014daa}, in scalar-torsion theories
\cite{Geng:2011aj,Hohmann:2018rwf}, etc.
$f(T)$ gravity proves to have interesting cosmological applications, being
efficient in describing both the late-time acceleration and the inflationary
phase, while its confrontation with observations leads to very satisfactory
results
\cite{Chen:2010va,Zheng:2010am,
Bamba:2010wb, Li:2011rn,
Capozziello:2011hj, Wu:2011kh, Wei:2011aa, Amoros:2013nxa,
Otalora:2013dsa,Bamba:2013jqa,
Li:2013xea,Ong:2013qja,Paliathanasis:2014iva,Nashed:2014lva,Darabi:2014dla,
Malekjani:2016mtm,
Farrugia:2016qqe,
Qi:2017xzl,
Bahamonde:2017wwk,Karpathopoulos:2017arc,Abedi:2018lkr,
Krssak:2018ywd,
Iosifidis:2018zwo, El-Zant:2018bsc, Anagnostopoulos:2019miu,Nunes:2019bjq,
Yan:2019gbw,ElHanafy:2019zhr,Saridakis:2019qwt,Wang:2020zfv,Bahamonde:2020lsm,
Briffa:2020qli, Hashim:2020sez}.
Notably, it has been shown that reconciling Planck with the local value of $H_0$ 
in a six-parameter space is achievable within the exponential infrared $f(T)$ 
gravity \cite{Hashim:2021pkq}

On the other hand, since the standard inflationary Big Bang scenario faces
 the crucial problem of the initial singularity (unavoidable in the case where
inflation is    realized using a scalar field in the framework of general
relativity \cite{Borde:1993xh}), a potential solution in terms of bouncing 
cosmologies has been introduced by considering  
Friedmann-Lema\^itre-Robertson-Walker (FLRW) models with a positive spatial 
curvature, where the matter sector is dominated by a massive scalar field 
\cite{Starobinskii:1978} (see also \cite{Mukhanov:1991zn,Novello:2008ra}). Since 
the
bounce realization requires the violation of the null energy condition, it can
be easily obtained in various modified gravity theories, such as  the Pre-Big-Bang
\cite{Veneziano:1991ek}
and the Ekpyrotic \cite{Khoury:2001wf,Khoury:2001bz} models,   higher-order
corrected gravity \cite{Brustein:1997cv,Tirtho1,Nojiri:2013ru}, even with the 
earlier work of $f(R)$ gravity \cite{Starobinsky:1980te} as well as other 
$f(R)$ models
\cite{Bamba:2013fha,Nojiri:2014zqa}, braneworld scenarios
\cite{Shtanov:2002mb,Saridakis:2007cf},
non-relativistic gravity \cite{Cai:2009in}, massive gravity
\cite{Cai:2012ag}, loop quantum
cosmology
\cite{Ashtekar:2006wn,Bojowald:2001xe,Ashtekar:2007em}, Finsler gravity
\cite{Minas:2019urp}  etc, while it  can be
easily obtained within $f(T)$ gravity too \cite{Cai:2011tc}. Although 
trans-Planckian problems, which may arise in inflationary models due to 
dispersion law modifications for frequencies beyond the Planck scale, have been 
discussed in Ref.~\cite{Martin:2000xs}, it has been shown that these problems do 
not exist  within inflationary scenarios as far as local Lorentz invariance is 
not broken even for those ultrahigh frequencies  
\cite{Starobinsky:2001kn,Starobinsky:2002rp}. Bounce cosmology proves efficient 
too to avoid these problems and bypasses the initial singularity 
\cite{Martin:2000xs,Brandenberger:2012aj}, while at the perturbation
level it leads to scale-invariant power spectrum 
\cite{Starobinsky:1979ty,Wands:1998yp,Finelli:2001sr,Biswas:2015kha}.

Nevertheless, the interesting question that appears is whether one can obtain a
unified description of the whole universe evolution through modified gravity.
In the classes where the initial singularity is bypassed through the bounce
realization, this question includes the investigation of time intervals up
to infinitely early times. On the other hand, in every cosmological scenario it
is always interesting and necessary to study the fate of the universe in the
asymptotically far future.

In the present work we are interested in investigating the complete universe
evolution in the framework of $f(T)$ cosmology. In particular, we desire to
construct a scenario that includes the standard observed thermal history of the
universe, namely the sequence of radiation, matter and late-time acceleration
epochs, and moreover bypasses the initial singularity. In order to achieve this
we take advantage of the fact that the field equations of $f(T)$ gravity are
second-ordered, and thus the corresponding autonomous dynamical  system
is  one dimensional
\cite{Bamba:2016gbu,ElHanafy:2017xsm,ElHanafy:2017sih,Awad:2017yod}. Hence, the
resulting phase space can be systematically explored, while still being much
more
complex than the corresponding one of general relativity, thus allowing for
significantly richer cosmological behavior.

 We organize this manuscript as follows. In Section \ref{Sec:2}, we examine the
necessary requirements on the kinematic level and we introduce  a
non-singular scale factor that can produce the unified universe evolution.
 In Section \ref{Sec:3}  we analyze in detail the resulting cosmology,
which we name   ``Pseudo-Bang  Scenario'', since its first phase presents the
features of a  Pseudo-Rip but in a reversed way. As we show,   the universe
passes through a phantom crossing, turnaround, bounce,
inflation, radiation, matter and late-time acceleration eras, and
asymptotically it results to a  Pseudo-Rip phase.
In Section \ref{Sec:4}  we reconstruct  the $f(T)$ theory which can
dynamically generate the above phase-portrait behavior, namely the unified
 Pseudo-Bang cosmological scenario. Additionally, we apply the
energy conditions and the inertial-force approach, which determine the ripping
behaviour, to verify that these phases fall within this classification.
Finally, in Section \ref{Sec:5}  we summarize the obtained results.

\section{Unifying bounce and late-time accelerated cosmologies}
\label{Sec:2}

In this section we investigate the kinematics of a cosmological scenario that
unifies the bouncing behavior with the standard thermal history of the universe
and in particular with the late-time accelerated era. In particular, we desire
to examine the necessary form of the Hubble parameter evolution $H(t)$ that is
required   to obtain  the aforementioned unified evolution.
In order to achieve this we apply the    dynamical system
approach and we focus our discussion to   one-dimensional autonomous systems,
i.e. where the Hubble derivative satisfies $\dot{H}=\mathcal{F}(H)$ (as we will
later see this is the case of $f(T)$ gravity).  
In this case, the 
differential equation represents a vector field on a \textit{line} ($H$-axis). 
We just need to draw the graph of $\mathcal{F}(H)$ and then use it to sketch the 
vector field on the real line (the $H$-axis). Therefore, it is still convenient 
to depict $\dot{H}$ versus $H$, and then insert arrows on the $H$-axis to 
indicate the corresponding Hubble flow vector at each $H$ in a simple way. The 
arrows point to the right when $\dot{H}> 0$, to the left when $\dot{H} < 0$, 
while for $\dot{H}=0$ there is no flow. Additionally, concerning the 
continuity and differentiability of $\mathcal{F}(H)$, given an initial 
condition $H(t_i) = H_i$, the continuity of $\mathcal{F}(H)$ guarantees the 
existence of a solution, while its differentiability guarantees the uniqueness 
of the solution (for more details see \cite{book:Steven}). This approach 
allows for the visualization of all possible cosmological solutions as a 
graphical representation of the phase portrait, independently of the initial 
conditions, where every phase point can serve as an initial condition.
\begin{figure}[h!]
\begin{center}
\includegraphics[scale=0.5]{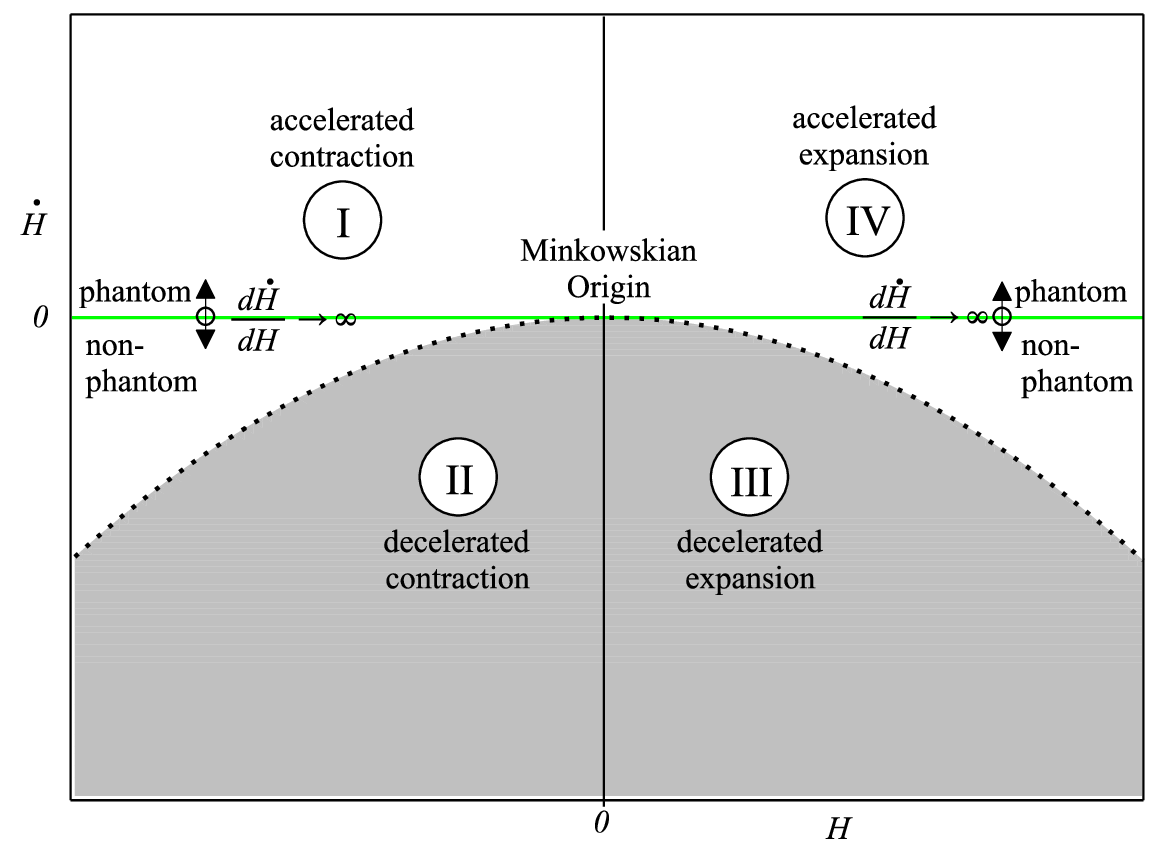}
\caption{{\it{Schematic representation of the structure of the
phase space diagram in the ($H$, $\dot{H}$) plane, and the classification of
the corresponding kinematic regions.}}}
\label{Fig:phasespace}
\end{center}
\end{figure}

In  Fig.~\ref{Fig:phasespace} we schematically present the structure of the
phase space diagram in the ($H$, $\dot{H}$) plane.
In general, the   phase space has a Minkowskian origin at ($0$, $0$). We define
the zero acceleration curve (dotted curve), which corresponds
to $q\equiv -1-\dot{H}/H^2=0$, and acts as a boundary between accelerated and
decelerated regions. We split the phase space into four kinematic regions
according to the values of $H$ and $q$ in each region:
The non-shaded region (I)
represents an accelerated contraction, since $H<0$ and $q<0$. The shaded region
(II) represents a decelerated contraction, since $H<0$ and $q>0$. The shaded
region (III) represents a decelerated expansion, since $H>0$ and $q>0$. The
non-shaded region (IV) represents an accelerated expansion, since $H>0$ and
$q<0$. We mention that the positive (negative) $\dot{H}$ leads to  phantom
(non-phantom) cosmology, respectively. The transition from phantom to
non-phantom or vise versa is allowed only through particular type of fixed points, characterized by infinite slope $d\dot{H}/dH \to \pm \infty$, which can be reached in finite time.

Having the above discussion in mind we deduce that  it is easy to
recognize complicated cosmological scenarios by following their phase
trajectories and studying their qualitative behaviours (see \cite{Awad:2017yod}
for more details). In the following subsections we present some specific
evolution behaviors.

\subsection{Standard bounce}\label{Sec:2.1}

Let us first investigate the conditions for the standard non-singular bounce
realization.
As it is known, this can be generated by employing a scale factor of the form
\cite{Novello:2008ra}
\begin{equation}
\label{bounce-scfac}
a(t)=a_{B}\left[\frac{3}{2}\gamma\alpha
(t-t_{B})^{2}+1\right]^{\frac{1}{3\gamma}},
\end{equation}
where the constant $a_{B}\equiv a(t_{B})$ is the minimal scale factor at the
bounce point $t_{B}$ and $\alpha$ is a positive   parameter with
dimensions [T]$^{-2}$. Moreover, $\gamma$ is the barotropic index related to
the equation-of-state parameter of the cosmic fluid as
\begin{equation}
\label{linear-EoS}
\gamma-1=w=p/\rho,
\end{equation}
where $p$ and $\rho$ are the pressure and  energy density respectively. Hence,
for positive $\gamma,\alpha$ the above scale factor is indeed non-singular for
finite times.
As one can see, the scale
factor (\ref{bounce-scfac}) generates a symmetric phase portrait about $H=0$ axis
\cite{ElHanafy:2017sih,Awad:2017yod}
\begin{equation}\label{st-bounce-PhasePortrait}
\dot{H}_{\pm}=\frac{3\gamma H^2 \sqrt{\alpha^2-6\alpha \gamma H^2}}{\pm \alpha
- \sqrt{\alpha^2-6\alpha \gamma H^2}},
\end{equation}
where $\dot{H}_{+}$ ($\dot{H}_{-}$) denotes the branch $\dot{H}>0$
($\dot{H}<0$).

\begin{figure}
\centering
\subfigure[~Bounce phase portrait]{\label{fig:st-bounce-phport}\includegraphics[scale=.37]
{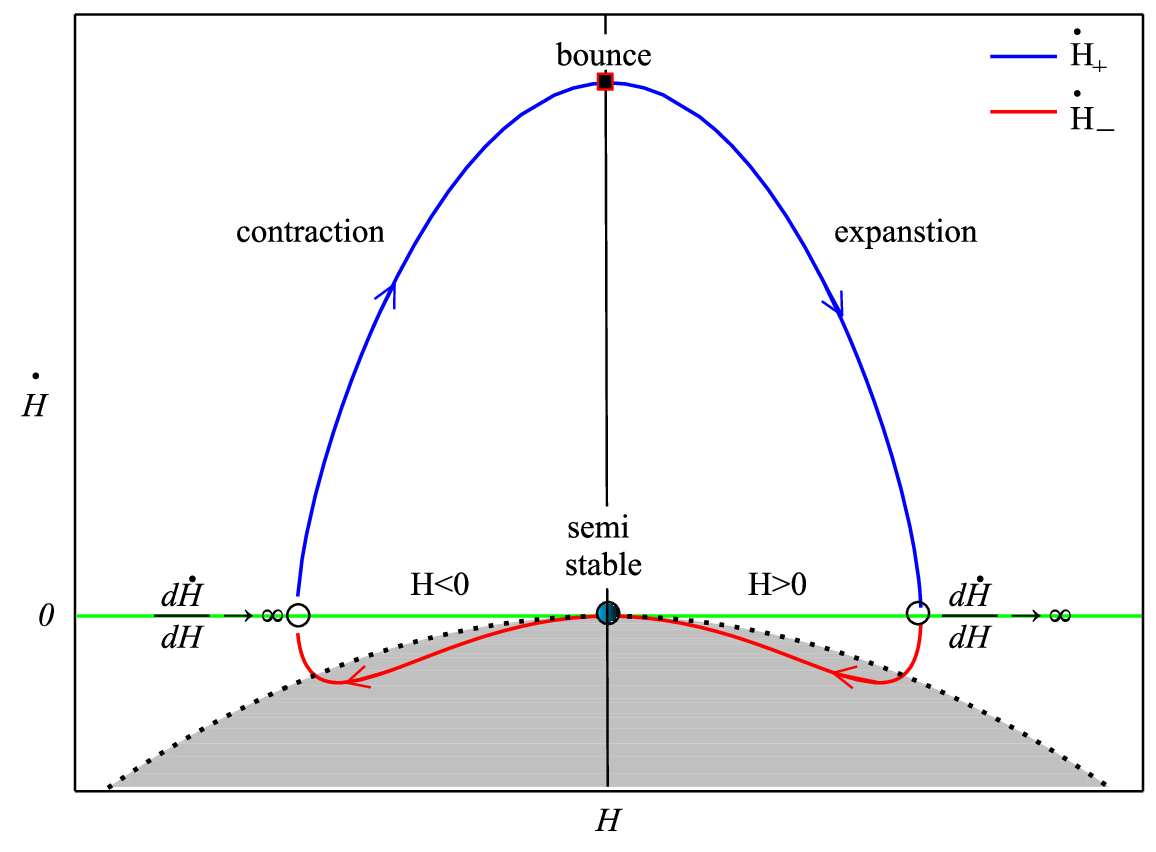}}\hspace{.7cm}
\subfigure[~Pseudo-Bang phase portrait]{\label{fig:unified-phport}\includegraphics[scale=.37]
{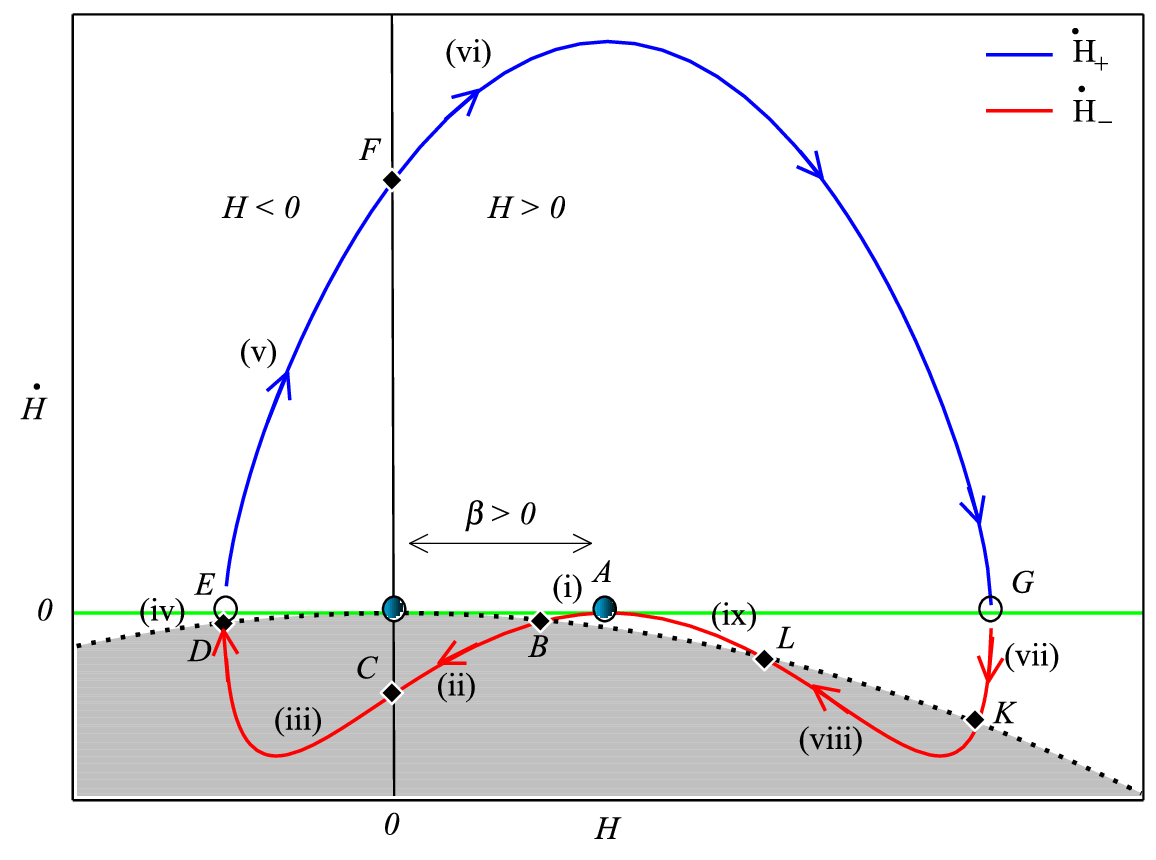}}
\caption[figtopcap]{\subref{fig:st-bounce-phport}
{\it{Schematic phase portrait
of the standard bounce cosmology
(\ref{st-bounce-PhasePortrait}). It has a reflection
symmetry around $H=0$, where the universe interpolates between two Minkowskian
phases.}}
\subref{fig:unified-phport}
{\it{Schematic phase portrait of the
Pseudo-Bang scenario (\ref{unified-PhasePortrait}). The universe has a
Pseudo-Bang origin, evolves in a first inflationary phase, a
cosmological turnaround and then   a bounce, after which we have a
second inflationary regime with a successful exit. Subsequently the universe
follows the standard thermal history of the sequence of radiation, matter and
late-time acceleration, resulting finally in an everlasting Pseudo-Rip phase
(see text).}}}
\label{Fig:unified-bounce}
\end{figure}

The non-singular bouncing cosmology phase portrait
(\ref{st-bounce-PhasePortrait}) is characterized by a double valued function,
as presented in Fig. \ref{Fig:unified-bounce}\subref{fig:st-bounce-phport}.
 Since the universe cannot reach the
Minkowskian origin at a finite time, the
universe is non-singular. We mention that the bounce occurs at $H=0$ where
$\dot{H}$ is positive. In fact, the $\dot{H}>0$ regions require an effectively
phantom cosmology. However, the crossing between phantom and non-phantom phase
is possible only wherever the phase portrait is double-valued and has vertical
slope at the crossing points (fixed points)\footnote{The conditions to reach a fixed point  in a finite time has been discussed in
\cite{Awad:2013tha} (also see \cite{Awad:2017yod}).}. Furthermore,  it is obvious
that the universe cannot result to a late-time accelerated expansion phase.
This
can be clearly seen in Fig.
\ref{Fig:unified-bounce}\subref{fig:st-bounce-phport}, since the last phase of
the portrait on the $\dot{H}_{-}$ branch remains in the shaded region III
eternally \cite{Awad:2017yod,ElHanafy:2017xsm}.

\subsection{Merging bounce with late-time acceleration}\label{Sec:2.2}

Let us try to   modify  the scale factor (\ref{bounce-scfac}) to additionally
obtain late-time accelerated expansion. One could think to  impose a
positive cosmological constant $\Lambda$  similarly to $\Lambda$CDM cosmology.
This would impose a vertical shift of the phase portrait of Fig.
\ref{Fig:unified-bounce}\subref{fig:st-bounce-phport} slightly upwards. Thus,
the universe begins at a de Sitter phase with negative $H=-H_{de}$, evolving
towards another de Sitter phase with positive $H=+H_{de}$ instead of the
Minkowski phase at $H=0$. However, the new feature is that the fixed points
at the left and right boundaries of the portrait, i.e. at $H=\pm H_{max}$ where
$H_{max}$ is the maximum value of the Hubble parameter, will not have infinite
slopes anymore, and therefore the transition between phantom and non-phantom
regimes, namely between $\dot{H}_{+}$ and $\dot{H}_{-}$ branches, cannot occur
in a finite time. Thus, we need to find an alternative way to     unify bounce
and late-time acceleration. This is done in the following.

Since adding by hand a positive   constant is not efficient, we proceed
  modifying the standard bounce scale factor (\ref{bounce-scfac}) by
introducing a correction exponential function as
\begin{equation}\label{unified-scfac}
a(t)=a_{k}~e^{\beta (t-t_{i})}\left[\frac{3}{2}\gamma\alpha
(t-t_{i})^{2}+1\right]^{\frac{1}{3\gamma}},
\end{equation}
where $a_{k}$ and $t_{i}$ are constants and $\beta$ is a positive
dimension-full parameter with dimensions [T]$^{-1}$, whereas  the
modified scale factor \eqref{unified-scfac} reduces to the usual
bouncing model by setting $\beta=0$. Unlike the bounce scale factor which has one local minimum at bounce $t=t_B$, the modified scale factor has a local minimum and local maximum at $t=t_i-\frac{1}{3\beta \gamma}\pm \frac{\sqrt{\alpha(\alpha-6\beta^2\gamma)}}{\alpha \beta \gamma}$, which implies that $\alpha>6\beta^2\gamma$. In fact, these local critical points are associated with turnaround and bounce phases.
The Hubble parameter corresponding to (\ref{unified-scfac}) is
\begin{equation}\label{unified-Hubble}
H(t)=\beta+\frac{2\alpha(t-t_{i})}{2+3\gamma \alpha (t-t_{i})^2},
\end{equation}
and its first derivative reads
\begin{equation}\label{unified-Hubble-dr}
\dot{H}(t)=
\frac{2\alpha\left[2-3\gamma\alpha(t-t_{i})^{2}\right]}{3\left[
2+3\gamma\alpha(t-t_ { i } )^ { 2 } \right]^{2}}.
\end{equation}
Note that these expressions remain finite for positive $\gamma,\alpha$, i.e in
the case where the bounce is non-singular. Inverting the above expression we
acquire
\begin{equation}\label{unified-time}
t_{\pm}(H)=t_i+\frac{\alpha\pm  \sqrt{\alpha^2-6 \gamma \alpha (H-\beta)^2}}{3
\gamma \alpha (H-\beta)},
\end{equation}
and thus inserting into (\ref{unified-Hubble-dr}) we obtain
  the   phase portrait equation
\begin{equation}\label{unified-PhasePortrait}
\dot{H}_{\pm}(H)=\frac{3\gamma (H-\beta)^2 \sqrt{\alpha^2-6\gamma \alpha
(H-\beta)^2}}{\pm \alpha - \sqrt{\alpha^2-6\gamma \alpha (H-\beta)^2}}.
\end{equation}
Note that  $H=\beta$ is a
fixed point and the time required to reach it is infinite.

The phase portrait graph of (\ref{unified-PhasePortrait}) is given in Fig.
\ref{Fig:unified-bounce}\subref{fig:unified-phport}, which indeed shows that
$H$ and $\dot{H}$ values are always finite. Moreover, we obtain   fixed
points, namely having $\dot{H}=0$, at $H=\beta$ (de Sitter origin) and at
maximum positive and minimum negative values of the Hubble parameter
$H=\pm\sqrt{\frac{\alpha}{6\gamma}}+\beta$, respectively. We note that in a viable scenario, one needs $|H|_\textrm{max} \gg \beta$ and subsequently $\alpha \gg 6 \beta^2 \gamma$. As it is clear from
the graph,  the modified scale factor (\ref{unified-scfac}) shifts the phase
portrait symmetry from line $H=0$ to $H=\beta$. As a result, the Minkowskian
origin of the conventional bounce is shifted to a de Sitter one, however this
has   crucial consequences on the cosmic evolution as we will discuss   in
Sec. \ref{Sec:3}. We mention that in this schematic graphical representation
we have exaggerated the value of the parameter $\beta$ in Fig.
\ref{Fig:unified-bounce}\subref{fig:unified-phport}, in order to show clearly
the breaking of the symmetry around $H=0$. However, as we will see later on,
when confronting with observational
data the value
of $\beta$ will be a small positive number, and the deformation seen in
\ref{Fig:unified-bounce}\subref{fig:unified-phport} will be small, but still
effective.

In the final phase of
the portrait of Fig. \ref{Fig:unified-bounce}\subref{fig:unified-phport}, the
universe
evolves towards a de Sitter fixed point at $H=\beta$, providing a late
accelerated
expansion phase. In addition, the fixed points at the phase portrait
boundaries, $H=\pm H_{max}$, are still having  infinite slopes which is a
necessary condition to allow for the crossing of the phantom divide   in
a smooth way
(a detailed
analysis of the phase portrait is given in Section \ref{Sec:3}).

In summary, up to now we have achieved our target to unify
bounce and late acceleration in a single scenario. The last step is to determine
  the model parameters in order to obtain a viable scenario.
 The above scenario   contains the four parameters $a_{k}$, $t_{i}$, $\alpha$
and $\beta$, in addition to the barotropic index (equation of state) parameter
$\gamma$, which takes the values $\gamma=1$ for dust matter and $\gamma=4/3$
for radiation. In order determine their values, one should thus choose four
conditions from the observed universe history.

In standard Big Bang cosmology there is the initial singularity at cosmic time
$t=0$. However, since in our case $-\infty < t < \infty$ it proves convenient
to set  $t=0$ at the fixed point $\dot{H}=0$ with $H_{max}=\sqrt{\frac{\alpha}{6\gamma}}+\beta$, which is identified with point
$G$ on the phase portrait of Fig.
\ref{Fig:unified-bounce}\subref{fig:unified-phport}. From now on all time identifications are calculated from the fixed point $G$. This choice enables us to
confront the model with the standard observational requirements of the cosmic thermal history. Imposing the fixed point phase condition $\dot{H}=0$ in Eq.
(\ref{unified-Hubble-dr}) and   solving for the cosmic time $t$  we acquire
\begin{equation}\label{dS-time}
    t^{\pm}=t_{i}\pm \sqrt{\frac{2}{3\gamma\alpha}},
\end{equation}
where $t^{-}$ ($t^{+}$) identifies the fixed point phase at $H<0$ ($H>0$) regime
at point $E$ ($G$) of the phase portrait Fig.
\ref{Fig:unified-bounce}\subref{fig:unified-phport}. Setting $t^{+}=0$ as
illustrated above,  we determine that
\begin{equation}\label{ti}
    t_{i}=-\sqrt{\frac{2}{3\gamma\alpha}}.
\end{equation}
Therefore, we find that  $t^{-}=2t_{i}=-2\sqrt{\frac{2}{3\gamma\alpha}}$ at
point $G$. This is the first out of the four conditions.
For the  other three we choose:\\
(i) For the present time we choose
$t_{0}\approx 4.3\times 10^{17}$ s as it arises from standard cosmology, and we
moreover normalize the present
  scale factor     to
$a_{0}=a(t_{0})=1$.\\
(ii) At the end of inflation (point $K$ on Fig.
\ref{Fig:unified-bounce}\subref{fig:unified-phport}), which corresponds to zero
  acceleration $\ddot{a}=0$ we need to have  $t\sim 10^{-32}$ s as expected
from standard cosmology.\\
(iii) For the late-time transition from deceleration to
acceleration (point $L$), where
the acceleration is again zero, namely $\ddot{a}=0$, we impose a time
$t\sim2.4\times 10^{17}$ s, which is consistent with the observed transition
redshift $z_{tr}\sim0.6-0.8$ \cite{Capozziello:2015rda}.
Finally, alongside conditions (i) and (iii)  we consider $\gamma=1$
as cold dark matter is expected to dominate the evolution, while for condition
(ii) we impose $\gamma=4/3$ as radiation is expected to be dominant at the
reheating phase by the end of inflation.
Hence, we conclude that
\begin{equation}
\label{parameters}
a_{K} \sim 1.2\times 10^{-33},\quad \alpha \sim 8.6\times
10^{62}~\textmd{s}^{-2},\quad \beta \sim 6.2\times 10^{-19}~\textmd{s}^{-1},
\end{equation}
while inserting into  (\ref{ti})  we find
\begin{equation}\label{ti1}
    t_{i}\sim-2.4\times 10^{-32}~\textmd{s}.
\end{equation}
The numerical results show that $\alpha \gg 6\beta^2 \gamma$, confirming 
the viability condition in order to have $|H|_\textrm{max} \gg \beta$, and also 
  verifying that the modified scale factor \eqref{unified-scfac} should have 
not only a local minimum as in the bounce scenario but also a local maximum 
which is associated with a turnaround phase. In the following we revisit these 
phases among other interesting features in more detail by investigating the 
corresponding phase portrait.

\section{The Pseudo-Bang  Scenario}
\label{Sec:3}

In this section  we utilize the  phase portrait analysis in order to study the
entire cosmic evolution and the stability of the scenario. As we described
above, introducing the parameter $\beta$ in the scale factor
results to a  symmetry shift in the phase portrait, and the line of symmetry
moves from Minkowski origin ($H=0$, $\dot{H}=0$) to de Sitter origin
($H=\beta$, $\dot{H}=0$).
This non-trivial shift seen in Fig.
\ref{Fig:unified-bounce}\subref{fig:unified-phport}, allows the phase portrait
to cut the $H=0$ line in a  non-trivial way twice. One of the intersection
points is as usual at the bounce point $F$ where $\dot{H}>0$, while the other
is
at the turnaround point $C$ where $\dot{H}<0$.

Let us present briefly the key points of the phase portrait of Fig.
\ref{Fig:unified-bounce}\subref{fig:unified-phport}. Point $A$ represents
the    de Sitter  phase, that is the eternal    phase of the
universe as $t\to \mp \infty$. Points $B$, $D$, $K$ and $L$ represent
transitions between acceleration and deceleration, which are characterized by
$\ddot{a}=0$. Points $C$ and $F$ represent the turnaround and
bouncing points, which are  characterized by $H=0$ with $\dot{H}<0$
and
$\dot{H}>0$, respectively. Points $E$ and $G$ represent a particular type of fixed points
($\dot{H}=0$) that can be exceptionally reached in finite time. As mentioned above,
this configuration allows the universe to cross the phantom divide line
smoothly. According to the numerical values of the model parameters
(\ref{parameters}), we summarize the results in Table \ref{T0}, estimating the
representative values of the scale factor, Hubble parameter, and the energy
scale $E\sim \sqrt{M_{p}H}$ at each   point, mentioning the corresponding
cosmological
features.
\begin{table}
\begin{center}\begin{tabular}{cccccc}
\hline\hline
Point&$t$(sec.)&$a$&$H$(GeV)&$E$(GeV)&cosmological phase\\
\hline
$A$ & $-\infty$ & $0$ & $4.1\times 10^{-43}$ & $  9.9\times 10^{-13}$ &
Pseudo-Bang \\
$B$ & $-1.9\times 10^{18}$ & $3.3\times 10^{-9}$ & $2.4\times 10^{-43}$ &
$7.6\times 10^{-13}$ & transition I\\
$C$ & $-8.0\times 10^{17}$ & $4.3\times 10^{-9}$ & $\approx 0$ & $\approx 0$
& turnaround\\
$D$ & $-5.8\times 10^{-32}$ & $1.6\times 10^{-33}$ & $-6.4\times 10^{6}$ &
$3.9\times 10^{12}$ & transition II\\
$E$ & $-4.8\times 10^{-32}$ & $1.5\times 10^{-33}$ & $-6.8\times 10^{6}$ &
$4.0\times 10^{12}$ & phantom crossing I\\
$F$ & $-2.4\times 10^{-32}$ & $1.2\times 10^{-33}$ & $\approx 0$ & $\approx
0$ & bounce\\
$G$ & $0$ & $1.5\times 10^{-33}$ & $6.8\times 10^{6}$ & $4.0\times 10^{12}$
& phantom crossing II\\
$K$ & $1.0\times 10^{-32}$ & $1.6\times 10^{-33}$ & $6.4\times 10^{6}$ &
$3.9\times 10^{12}$ & transition III\\
$L$ & $2.4\times 10^{17}$ & $0.6$ & $2.2\times 10^{-42}$ &
$2.3\times 10^{-12}$ & transition IV\\
$A$ & $+\infty$ & $+\infty$ & $4.1\times 10^{-43}$ & $9.9\times 10^{-13}$ &
Pseudo-Rip\\
\hline
\end{tabular}\end{center}
\caption{Approximate estimations of the Hubble and the Energy scale at
the key points of the phase portrait of Fig.
\ref{Fig:unified-bounce}\subref{fig:unified-phport}, using the parameter values
(\ref{parameters}).  We refer to the cosmological phase transition I (at $B$)
as
a transition from accelerated to decelerated expansion, transition II (at $C$)
as a transition from decelerated contraction to accelerated contraction,
transition III (at $K$) as a transition from accelerated expansion to
decelerated expansion and transition IV (at $L$) as a late-time transition from
decelerated  to accelerated expansion. All quantities at different phases are calculated for radiation except at the last two phases $L$ and $A$ we take $\gamma = 1$ as matter becomes dominating.}\label{T0}
\end{table}

In the following subsections we discuss the features of each point and its
corresponding phase in more details.

\subsection{From Pseudo-Bang origin to Pseudo-Rip fate}
\label{Sec:3.1x}
\subsection*{Pseudo-Bang origin}
\label{Sec:3.1}

According to the phase portrait the cosmic  time flows clockwise, where the
origin has been shifted from Minkowski to a semi-stable de Sitter fixed point
$A$, at which $H=\beta\sim 4.1\times 10^{-43}$ GeV and $\dot{H}=0$. Using the
phase portrait equation (\ref{unified-PhasePortrait}), the flow time from
any phase point $H_{0}\lesssim \beta$ on the $\dot{H}_{-}$ branch can be
calculated by
\begin{equation}\label{time}
t=\int_{H_{0}}^{\beta}\frac{dH}{\dot{H}_{-}}\to -\infty,
\end{equation}
since $1\leq \gamma \leq 2$ to maintain the stability  and the causality
conditions. Therefore, the universe is eternal and has no initial finite-time
singularity. It is straightforward to show that the time asymptotic of the
scale factor (\ref{unified-scfac}) and the Hubble parameter
(\ref{unified-Hubble}) are respectively
\begin{equation}
\label{Pseudo-Bang -condition}
    \lim_{t\to -\infty} a(t)=0, \quad \lim_{t\to -\infty} H(t)=\beta^{-},
\end{equation}
given that $\beta>0$ and $\alpha>0$.

We mention that  in the standard bounce
the scale factor diverges as $t\to -\infty$. On the other hand, in standard Big
Bang cosmology the initial scale factor $a(t)\to 0$ and $H(t)\to \infty$ as
$t\to 0$. On the contrary, in the pure cosmological constant universe, the
Hubble parameter has a finite constant value, but the universe cannot exhibit
 a decelerated expansion phase. Hence, one can realize that the present
scenario is a novel one, in which the universe initial state  is
  intermediate between the Big Bang and the de Sitter universe. Inspired by
the Pseudo-Rip terminology (see below) we call
this eternal phase as \textit{Pseudo-Bang}, since it is characterized by
$a(t)\to 0$ and $H(t)\to const.>0$ as $t\to -\infty$ (note that this is
different from the emergent universe scenario which is characterized by
$a(t)\to
const.$ at $t\to -\infty$ \cite{Ellis:2002we,Mulryne:2005ef}). This is the
first
phase
of the scenario at hand.

\subsection*{Inflation I}\label{Sec:3.2}

Following the phase portrait  of Fig. \ref{Fig:unified-bounce}\subref{fig:unified-phport} clockwise, we identify the interval (i), which ends at point $B$, in which the portrait
cuts the zero acceleration curve $\dot{H}_{-}=-H^2$. Since the Hubble values are positive (as given in Table \ref{T0}, $2.4\times 10^{-43} \lesssim H(t) \lesssim 4.1\times 10^{-43}$ GeV) in this interval, and $\dot{H}_{-}>-H^2$, the universe expands with acceleration. Such an unconventional initial phase represents a non-singular inflationary phase, unlike standard bounce cosmology which begins with a decelerated contraction phase. This is the second phase of the scenario at hand.

\begin{figure}
\begin{center}
\includegraphics[width=\textwidth]{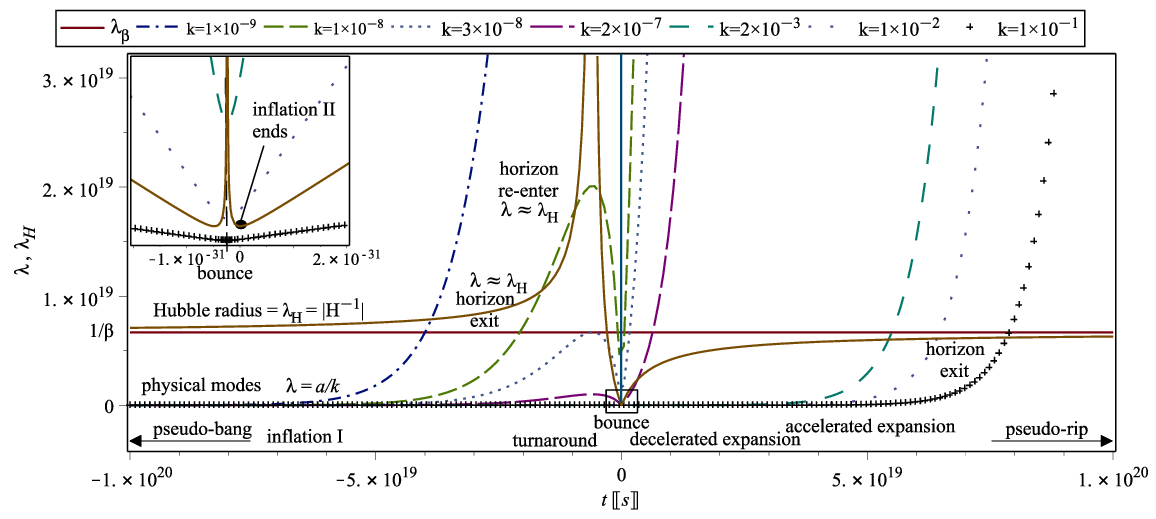}
\caption{
{\it{The time evolution of the wavelengthes $\lambda=a(t)/k$, where $a(t)$ is
given by Eq. (\ref{unified-scfac}), as well as the evolution of the Hubble
horizon $\lambda_{H}=|H^{-1}|$, where $H$ is given by Eq.
(\ref{unified-Hubble}).  The horizontal line identifies the minimal value of
the Hubble radius at Pseudo-Bang origin $\lambda_\beta \to 1/\beta$, as well as
its maximal value at Pseudo-Rip fate. The sub-graph   in the
upper left corner   zooms at  the wavelength and Hubble radius evolution
during the bounce phase and the end of inflation II, followed by a
decelerated expansion phase consistent with   standard cosmology. The model
parameters have been
set as in
(\ref{parameters}). It is clear that at the  Pseudo-Bang
all modes are sub-horizon, namely $\lambda \ll \lambda_\beta$. The long
wavelength modes (small $k\lesssim 10^{-8}$) exit the Hubble horizon by the end
of inflationary epoch I, while shorter wavelength modes ($k \gtrsim 10^{-3}$)
still remain sub-horizon. At the turnaround and bounce points $\lambda_H$
becomes infinite, since $H=0$, and all modes re-enter the horizon and become
sub-horizon. Subsequently, long
wavelength modes exit the horizon by the end of inflationary epoch II, while
  shorter wavelength modes  are sub-horizon and they
  exit the horizon at the later accelerated expansion phase.}}}
\label{Fig:horizon}
\end{center}
\end{figure}

Remarkably this quasi de Sitter inflationary phase runs at low energy level which in effect might contribute to solve the theoretical problem of the cosmological
constant. Additionally, since this phase is eternal, it has enough time to solve the usual problems of the Standard Model of cosmology. Furthermore, there is no need to compute the minimum $e$-fold number, $N=-\ln (a_i/a_f)$, since $a_i$ can always be chosen small enough to obtain a suitable $N$. On the other hand, in the scenario at hand the comoving Hubble radius $R_{H}=\frac{1}{a(t) |H(t)|}$, which we refer to as horizon, is infinite at the Pseudo-Bang  origin as $a(t)\to 0$. Therefore, all modes are at sub-horizon scale $k \gg R_{H}^{-1}$, or equivalently the physical wavelengths $\lambda \ll \lambda_{H}$, where $\lambda(t)=a(t)/k$ is the physical mode wavelength $\lambda_{H}=|H^{-1}|$ is the Hubble radius (we refer to the Hubble radius at Pseudo-Bang as $\lambda_\beta \to 1/\beta$). This is a necessary condition to have an initial causal universe. Consequently, we can assume that the primordial fluctuations are coherent, as indicated by  the observations of acoustic peaks in the power spectrum of Cosmic Microwave Background (CMB) anisotropies. In this case, it  is natural to assume that the quantum fluctuations around the initial vacuum state form the Bunch-Davies vacuum \cite{Bunch:1978yq}. The evolution of the physical modes at different scales verses Hubble horizon is given in Fig. \ref{Fig:horizon}.

\subsection*{Turnaround}\label{Sec:3.3}

At point $C$ on Fig. \ref{Fig:unified-bounce}\subref{fig:unified-phport} the universe goes from dynamical region III (decelerated expansion) to another dynamical region II (decelerated contraction). Therefore, point $C$
represents a turnaround point, $H=0$ and $\dot{H}_{-}<0$, at which the universe
reaches a maximum size with a finite deceleration unlike Big Brake models \cite{Keresztes:2012zn}. This is the third phase
of the scenario at hand.

At the turnaround point $C$,  the Hubble parameter goes to zero. Using
(\ref{unified-Hubble}), (\ref{parameters}) and (\ref{ti1}), we determine the
time at the turnaround as
\begin{equation}\label{tC}
    t_{C}=-\frac{1}{3\gamma\beta}-\sqrt{\frac{2}{3\gamma\alpha}}-
\frac{\sqrt{\alpha^2-6\gamma\alpha\beta^2}}{3\gamma\alpha\beta}\approx
-8\times 10^{17}~s.
\end{equation}
Notably, in general relativity, at turnaround phase one should introduce unconventional matter species violating the null energy condition. This situation can be avoided in modified gravity making the gravitational sector to compensate the matter component at the turnaround point \cite{Cai:2011bs}. Finally, we mention  that Fig. \ref{Fig:horizon} shows that  the corresponding
Hubble radius
$\lambda_H$ becomes infinite, allowing the modes to re-enter the horizon and
become sub-horizon again.

\subsection*{Phantom crossing I}\label{Sec:3.4}

At point $D$,
the phase portrait intersects the zero acceleration curve
$\dot{H}_{-}=-H^{2}$  for the second time, but at a negative value of $H\sim
-6.4\times 10^{6}$ GeV, as given in Table \ref{T0}. The, the universe enters in a new phase of accelerated contraction $\dot{H}_{-}> -H^{2}$, namely interval (iv). 
Notably, the point $E$ is a fixed point which exceptionally can be reached in finite time. This can be proven, since
\begin{equation}\label{TypeIVE}
\nonumber   \lim_{H\to H_E}\dot{H}_{\pm}= 0,\quad
\lim_{H\to H_E}\frac{d\dot{H}_{\pm}}{dH}= + \infty,\textmd{ and} \quad
t\to \int_{H<\beta}^{H_{E}}\dot{H}^{-1}_{\pm}dH = finite.
\end{equation}
Additionally the phase portrait is double valued about $E$ which is a necessary condition to cross the phantom divide line through point $E$, for more detail see \cite{Awad:2013tha,Awad:2017yod}. At this point $E$, we calculate $H_{E}=\beta-\frac{\sqrt{\alpha}}{3\gamma}\sim -6.8\times 10^{6}$ GeV, then the time to reach is $t_{E}=2t_{i}=-2\sqrt{\frac{2}{3\gamma\alpha}}\sim
-4.8\times 10^{-32}$ s.
This Phantom crossing I phase is the fourth epoch of the scenario at hand.

\subsection*{Bounce}\label{Sec:3.5}
At point $F$ the universe bounces from an accelerated contraction to accelerated expansion in an effective phantom regime, where $\dot{H}_{+}>0$, as it is shown in Fig.
\ref{Fig:unified-bounce}\subref{fig:unified-phport}. Using (\ref{unified-Hubble}) and (\ref{ti}), we determine the
time at the bounce point as
\begin{equation}\label{tF}
    t_{F}=-\frac{1}{3\gamma\beta}
-\sqrt{\frac{2}{3\gamma\alpha}}+\frac{\sqrt{\alpha^2-6\gamma\alpha\beta^2}}{
3\gamma\alpha\beta}\approx -2.4\times 10^{-32}~s.
\end{equation}
At the bounce point $F$ one should investigate two major problems that are usually facing such models in GR: The first is known as ghost instability due to violation of the null energy condition, which has been shown to be avoided within modified gravity framework whereas the matter sector remains casual and stable \cite{Cai:2011tc,Bamba:2016gbu}. The second is more severe and called the anisotropy problem since any small anisotropy evolves as $\propto a(t)^{-6}$ and becomes dominant during the contraction phase at suitably small scale factors, thus destroying FLRW
geometry \cite{Novello:2008ra}. Nevertheless, the usual way for its avoidance is the introduction of a super-stiff matter component $\gamma \gg 2$ which scales as $a(t)^{-3\gamma}$ and thus it grows faster and dominates over the anisotropy. This being the case as we will discuss this later on in Subsection \ref{Sec:4.4}. This bouncing phase is the fifth phase in the scenario at hand.
\subsection*{Phantom crossing II}\label{Sec:3.6}
Similar to the phase at the fixed point $E$, it can be shown that
\begin{equation}\label{TypeIVG}
\nonumber   \lim_{H\to H_G} \dot{H}_{\pm}= 0,\quad
\lim_{H\to H_G} \frac{d\dot{H}_{\pm}}{dH}= - \infty, \quad
\textmd{ and} \quad
t\to \int_{H>\beta}^{H_{G}}\dot{H}^{-1}_{\pm}dH = finite.
\end{equation}
Then, the fixed point $G$ can be reached in a finite time, whereas the portrait is double valued about it. Therefore, at the phase point $G$, the universe crosses the phantom divide line for the second time, similarly to point $E$, but this time from phantom to non-phantom phase. At that point the Hubble parameter $H_{G}=\beta+\frac{\sqrt{\alpha}}{3\gamma}\sim 6.8\times 10^{6}$ GeV. We stress
that we have set the time at point $G$ as $t_{G}=0$ as discussed in detail in Subsection \ref{Sec:2.2}. The Phantom crossing II is the sixth phase of the evolution.

\subsection*{Inflation II and graceful exit }\label{Sec:3.7}

During the interval (vii) the universe evolves into   quintessence
regime just as in standard inflationary models. In the latter, the
universe needs to enlarge itself around $10^{28}$ times ($\sim 60$ $e$-folds)
to solve the  horizon and flatness  problems of   standard
cosmology \cite{Olive:1989nu}. However, in our scenario there is no need for
this restriction, since the preceding phases are sufficient to solve   these
problems. Indeed, the model predicts just  few $e$-folds during this
accelerated expansion period as indicated in Table \ref{T0}. Additionally,
as the Table shows, the energy scales are $H\sim 10^{6}$ GeV and $E\sim
10^{12}$ GeV at the graceful exit point, that is below the energy scale of the
grand unified theory (GUT), and hence monopoles are not going to be produced,
thus bypassing the monopole problem of standard cosmology. Moreover, as it has
been shown,   the scale invariant power spectrum can be produced in
standard bouncing cosmology  in the contraction phase, which
occurs in the scenario at hand too. At the end of this interval, at
point $K$, the universe gracefully exits into a decelerated expansion phase,
which characterizes   standard cosmology. This smooth
transition period is essential to prepare the universe to begin the hot Big
Bang nucleosynthesis process. The graceful exit from inflation II is the
seventh phase of the present scenario.

\subsection*{Radiation and matter epochs}
\label{Sec:3.8}

The eighth phase of the universe evolution is the standard cosmology phase of
reheating, radiation and matter epochs. In this case,
it is convenient to examine the behaviour of
the phase portrait (\ref{unified-PhasePortrait}) at relevant epochs. In particular, for the
$\dot{H}_{-}$ branch at $H\gg \beta$, we obtain the leading term $\dot{H}_{-}=-\frac{3}{2}\gamma H^2$ which reproduces the
power-law standard cosmology phase portrait.

In order to obtain a fuller view and to examine the capability of the present scenario to predict a successful thermal evolution, we define the
entropy $S$ of all particles in thermal equilibrium at temperature $\Theta$ in
volume $V$. According to the first law of thermodynamics, in the expanding
universe  we have
\begin{equation}\label{Thermodynamics2}
    \Theta dS=d(\rho V)+p dV,
\end{equation}
with the integrability condition
$
    \frac{\partial^2 S}{\partial \Theta \partial V}=\frac{\partial^2
S}{\partial
V \partial \Theta}
$  \cite{Weinberg:1972kfs},
where
the energy density and pressure satisfy
\begin{equation}\label{Temp}
    \frac{dp}{d\Theta}=\frac{\rho+p}{\Theta} \Leftrightarrow
\frac{d\Theta}{\Theta}=\frac{\gamma-1}{\gamma}\frac{d\rho}{\rho}.
\end{equation}
From  the matter conservation equation
$    \dot{\rho}+3 \gamma H \rho=0$ it is implied that
$\rho\propto a^{-3\gamma}$. Solving (\ref{Temp})  we evaluate the temperature
as $\Theta(t)=\Theta_{0}~
a(t)^{-3(\gamma-1)}$, with $\Theta_{0}\equiv\Theta(t_{0})$   an arbitrary
constant. Hence, this finally leads to
\begin{equation}
\label{Temperature}
\Theta(t)=\Theta_{0}a_{k}^{-3(\gamma-1)}e^{-3(\gamma-1)\beta(t-t_{i})}
{\left[\frac{3}{2}\gamma\alpha(t-t_{i})^{2}+1\right]^{\frac{1-\gamma}{\gamma}}}.
\end{equation}
 We choose a boundary condition such that the temperature $\Theta\sim
2.73$ K at the present time $t_{0}\sim 10^{17}~\textmd{s}>t_{eq}$, with a dust
equation-of-state parameter $\gamma=1$. This determines the value $\Theta_{0}=
2.73$ K.

\begin{figure}[ht]
\begin{center}
\includegraphics[scale=0.5]{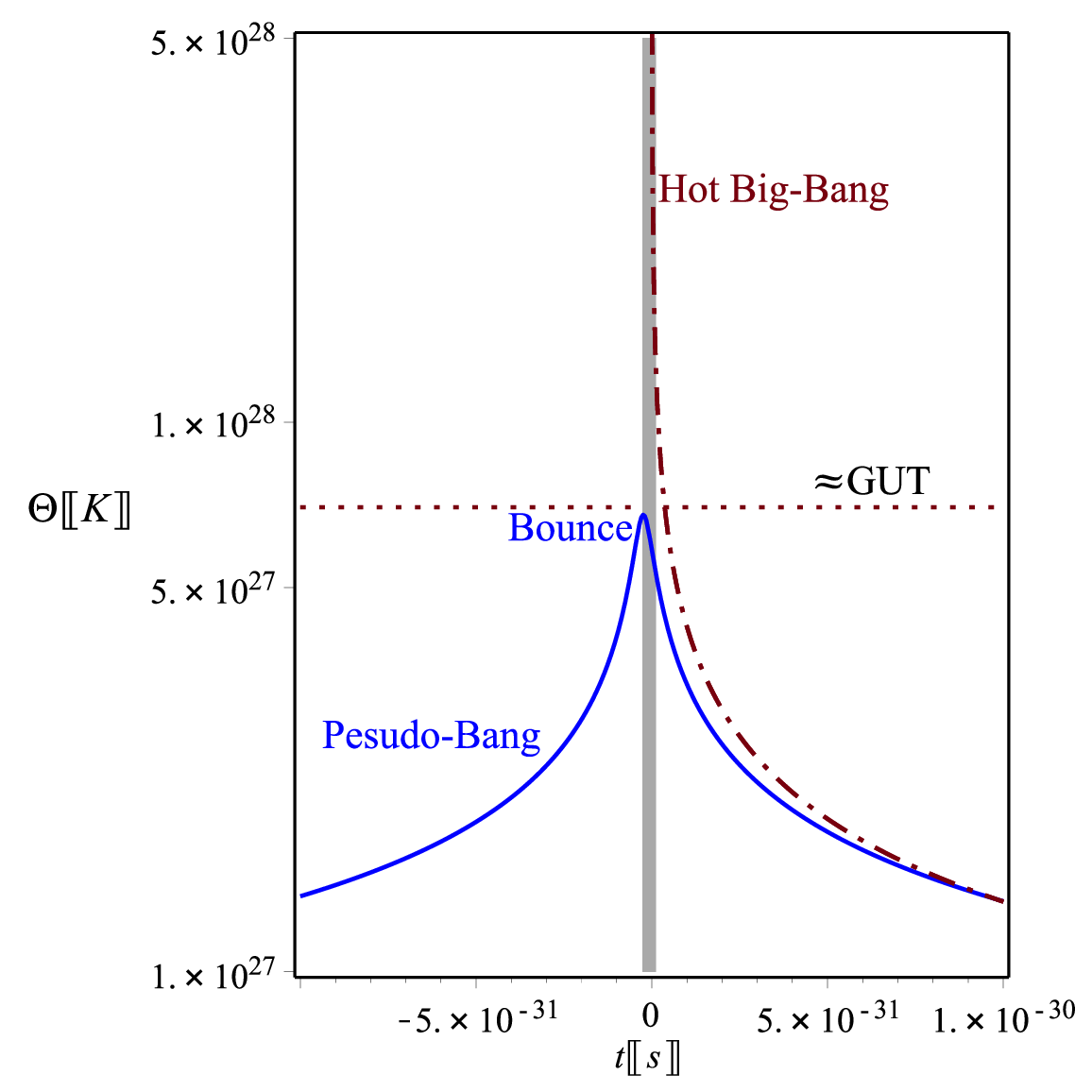}
\caption{
{\it{Temperature evolution is standard hot Big-Bang cosmology and in the
Pseudo-Ban scenario. In the former it  is characterized by
a
diverging temperature at the initial singularity $t=0$.
In the Pseudo-Bang  scenario  the temperature  (\ref{Temperature}) becomes
maximum at the bounce point, and then coincides   with the hot Big-Bang one.
Note that the bounce in the Pseudo-Bang  scenario does not occur at $t=0$ as
 in   standard bounce cosmology, and the shift form $t=0$ is indicated by
the grey region. The model parameters have been
set as in
(\ref{parameters}).}}}
\label{Fig:Temperature}
\end{center}
\end{figure}

In summary, at the contracting Pseudo-Bang   phase  the temperature increases
  and peaks up at the bounce point  with $\Theta\sim 10^{27}$ K (assuming
relativistic species with barotropic index $\gamma\approx 1.28$), and then it
 decreases following the standard thermal history.
In Fig. \ref{Fig:Temperature} we depict the
temperature evolution of the Pseudo-Bang  scenario on top of the corresponding
one
for the standard Big-Bang cosmology (we restrict  the graph to the interval
from Planck era to the radiation-matter equality, which is suitable for
comparison since at later times      both
scenarios exhibit the same behaviour, namely the same matter epoch.

\subsection*{Late-time acceleration}\label{Sec:3.9}

The ninth phase of the scenario at hand is the late-time acceleration. In
particular, on the phase portrait of Fig.
\ref{Fig:unified-bounce}\subref{fig:unified-phport} one can see that the
universe exhibits a late-time transition from deceleration to
acceleration, as indicated by point $L$. At this point, the phase portrait
crosses the zero acceleration curve for the last time, from
$\dot{H}_{+}<-H^{2}$ decelerated expansion region into $\dot{H}_{+}>-H^{2}$
accelerated expansion, where $H>0$ and $\dot{H}_{+}<0$. More precisely,
we determine the current value of the Hubble parameter as expected, by  taking
the present time $t_0=4.3\times 10^{17}$~s (which is consistent with the
assumption $a(t_0)=1$ or equivalently of redshift
$z(t_{0})=\frac{1}{a(t_0)}-1=0$) and according to the estimated values of the
model parameters (\ref{parameters}), (\ref{ti1}), and thus  the Hubble
function (\ref{unified-Hubble}) gives the present Hubble value $H_0\approx
66.6$ km/s/Mpc. Moreover, it is straightforward to show that at the
deceleration-to-acceleration transition time $t_L=2.4\times 10^{17}$~s the
corresponding redshift $z_{tr}=\frac{1}{a(t_L)}-1\approx 0.7$   lies  within
1$\gamma$ agreement with its measured value $z_{tr}=0.72\pm 0.05$ for
$H_0=68\pm
2.8$ km/s/Mpc \cite{Farooq:2016zwm}.
The late acceleration is labeled on the
phase portrait as interval (ix), and is confined between points $L$ and $A$.
Notably, this last phase cannot be exhibited  in the standard bounce of   Fig.
\ref{Fig:unified-bounce}\subref{fig:st-bounce-phport},  where the
universe continues the decelerated expansion eternally.

\subsection*{Pseudo-Rip fate}\label{Sec:3.10}

As we observe in Fig. \ref{Fig:unified-bounce}\subref{fig:unified-phport}, the
phase portrait
evolves towards the fixed point $A$ as a final fate. Similarly to
(\ref{time}),
the time needed to reach that point in infinite.
Given that $\beta>0$ and $\alpha>0$, the time asymptotic of the scale factor (\ref{unified-scfac}) and Hubble (\ref{unified-Hubble}) have the limits $ a(t)\to \infty$ and  $H(t) \to\beta^{+}$ as $t\to \infty$. Among four possible final phases of expanding universe classified in Ref.~\cite{Frampton:2011aa}, we deduce that in the scenario at hand the universe evolves towards a Pseudo-Rip which is an intermediate between the no Rip and the Little Rip. This phase is characterized by a monotonically increasing Hubble to a constant value, producing an
inertial force which does not increase monotonically but peaks up at a particular future time and then decreases. In this case the bound
structures dissociate if they are at or below a particular threshold that depends on the inertial force, for more details see Subsection \ref{Sec:4.4}. This is the tenth and final phase. 

\subsection{Poincar\'{e} patches}\label{Sec:3.11}
Recalling the Pseudo-Bang scale factor \eqref{unified-scfac}, one may realize that the hypersurface at the Pseudo-Bang $t=-\infty$ is not space-like but null. In this sense, the Pseudo-Bang resembles the de Sitter universe up to a sub-leading term, as it is clear from   \eqref{unified-Hubble}. Similarly to de Sitter, there can be   particles coming to Pseudo-Bang from the other part of spacetime. To reveal the similarity between the Pseudo-Bang phase and de Sitter universe, we impose the transformation
\begin{equation}
    \tau=\tilde{t}+\ln{\left[a_k^{1/\beta}\left(\frac{3}{2}\gamma \alpha \tilde{t}^2+1 \right)\right]},
\end{equation}
where $\tau\to \pm \infty$ as $\tilde{t}\equiv t-t_i\to \pm \infty$, which leads to the Pseudo-Bang scale factor as $a(\tau)=e^{\beta\tau}$. Moreover, we write
\begin{equation}
    \frac{d\tau}{d\tilde{t}}=1+\frac{3\gamma \alpha \tilde{t}}{\frac{3}{2}\gamma \alpha \tilde{t}^2+1},
\end{equation}
which implies that $d\tau$ differs from $d\tilde{t}$ by a sub-leading term only, and thus $d\tau/d\tilde{t}\to 1$ as $\tilde{t}\to \pm \infty$ ($\tau\to\pm\infty$). In this case, we write the FLRW metric at Pseudo-Bang/Rip as
\begin{equation}\label{dS_patch}
    ds^2=d\tau^2-e^{2\beta \tau} d\vec{x}~{^2}.
\end{equation}
The symmetry of the obtained spacetime can be revealed by making use of a flat embedding of \eqref{dS_patch} into a 5-dimensional Minkowski spacetime,
\begin{equation}
    ds_5^2=dX_0^2-dX_1^2-dX_2^2-dX_3^2-dX_4^2,
\end{equation}
which can be realized as the   hyperboloid 
\begin{equation}
    X_0^2-X_1^2-X_2^2-X_3^2-X_4^2=-\beta^{-2}.
\end{equation}
\begin{figure}
\centering
\subfigure[~Penrose–Carter diagram]{\label{fig:Penrose}\includegraphics[scale=.27]
{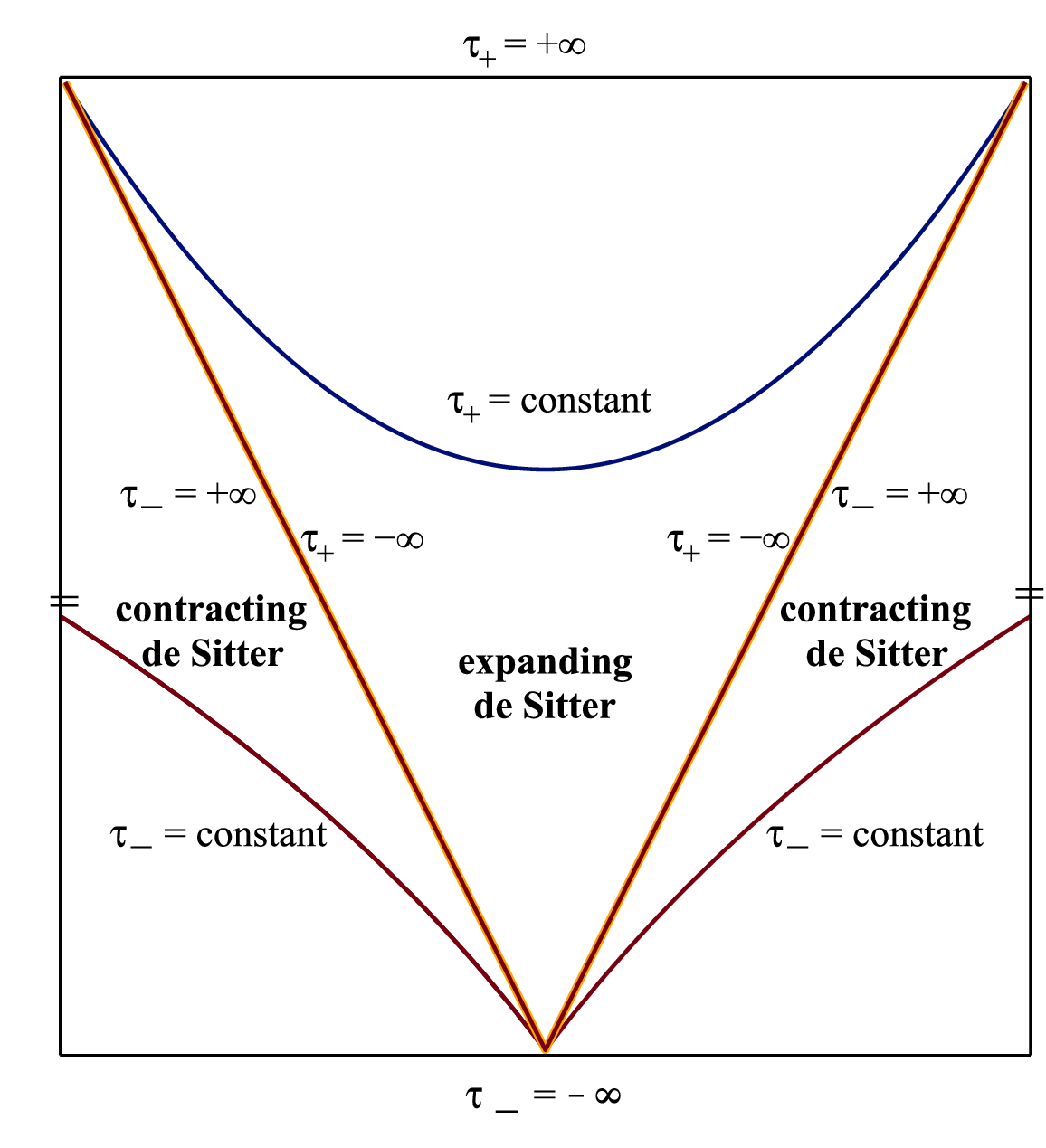}}\hspace{.7cm}
\subfigure[~Matching phase portraits]{\label{fig:matching}\includegraphics[scale=.48]
{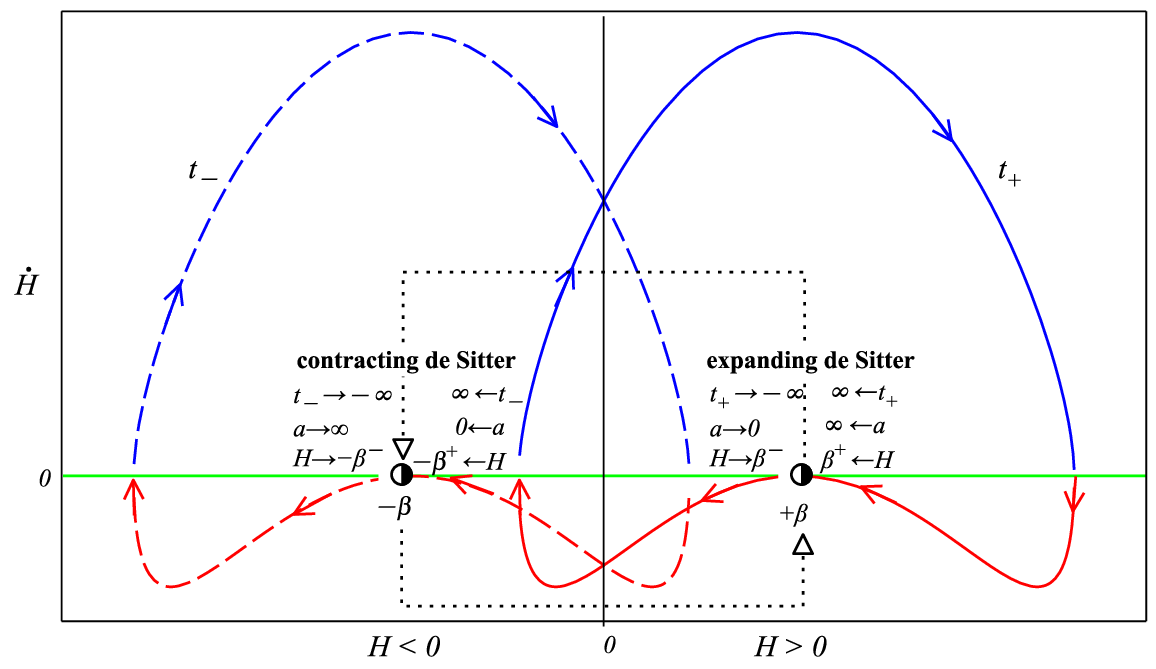}}
\caption[figtopcap]{\subref{fig:Penrose}
{\it{The entire Pseudo-Bang space where the expanding patch \eqref{exp_patch} is constrained by $-\infty < \tau_+ < \infty$ and the contracting patch \eqref{con_patch} is constrained by $-\infty<\tau_-<\infty$. The boundary $\tau_\pm=\mp \infty$ is light-like.}}
\subref{fig:matching}
{\it{The phase portrait is invariant under $\tilde{t}_+\to -\tilde{t}_-$ and $\beta\to -\beta$, and the matching between the two de Sitter semi-stable fixed points $H=\pm \beta$ is viable.}}}
\label{Fig:deSitter}
\end{figure}
For the coordinate transformation\footnote{We use $-\infty < \tau_+ < \infty$ to identify the expanding patch which represents only one-half of the entire de Sitter space, while $-\infty < \tau_- < \infty$ will be used to describe the other contracting patch.}
\begin{eqnarray}
\nonumber X_0&=&\frac{1}{\beta}\sinh{\left(\beta \tau_+\right)}+\frac{\beta}{2}e^{\beta \tau_+} ||\vec{x}||^2, \quad -\infty<\tau_+<\infty\\
\nonumber X_i&=&x^i_+ e^{\beta \tau_+}, \quad -\infty<x^i_+<\infty \,(i=1, 2, 3)\\
 X_4&=&\frac{1}{\beta}\cosh{\left(\beta \tau_+\right)}-\frac{\beta}{2}e^{\beta \tau_+} ||\Vec{x}||^2,\label{flat_coo}
\end{eqnarray}
the induced metric is
\begin{equation}\label{exp_patch}
    ds_+^2=d\tau_+^2-e^{2\beta \tau_+} d\Vec{x}{_+^2}.
\end{equation}
This shows that the above metric covers only half of the de Sitter spacetime, where Eqs. \eqref{flat_coo} restrict the spacetime configuration to the region $X_0+X_4=\beta^{-1}e^{\beta \tau_+}\geq 0$, that is the expanding Poincar\'{e} patch (EPP). The other half, $X_0+X_4\leq 0$, is covered by the metric
\begin{equation}\label{con_patch}
    ds_-^2=d\tau_-^2-e^{2\beta \tau_-} d\Vec{x}{_-^2},
\end{equation}
which represents the contracting Poincar\'{e} patch (CPP).

The entire manifold at Pseudo-Bang and Pseudo-Rip can be obtained via a Penrose–Carter diagram, as seen in Fig. \ref{Fig:deSitter}\subref{fig:Penrose}. This diagram reveals that the boundary between the EPP and the CPP at $\tau_\pm=\mp\infty$ is light-like. Note that the solution is invariant under the transformations $t_+\to -t_-$ and $\beta\to -\beta$. In terms of the phase-portrait terminology, one obtains that the phase portrait equation \eqref{unified-PhasePortrait} may have another fixed point of de Sitter type at $H=- \beta$. In Subsection \ref{Sec:3.1x}, we discussed the evolution corresponding to the expanding de Sitter origin, i.e. $H=+\beta$, as realized by the phase portrait of Fig. \ref{Fig:unified-bounce}\subref{fig:unified-phport}. Since the Pseudo-Bang resembles the de Sitter case, and $t\to -\infty$ is a null hypersurface, it may not represent a genuine beginning. For that reason we label the time of the portrait with the origin at $H=\beta$ as $-\infty< t_+< \infty$. On the other hand, we label the time of the portrait with origin at $H=-\beta$ as $-\infty < t_- < \infty$, as seen in Fig. \ref{Fig:deSitter}\subref{fig:matching}. One realizes that there is no need for junction conditions to match the two phases at $H=\pm\beta$. Indeed, the Pseudo-Bang in the expanding de Sitter (i.e. EPP), $t_+\to -\infty$, $a(t_+)\to 0$ and $H(t_+)\to \beta^{-}$, is naturally matched with the contracting de Sitter (i.e. CPP), $t_-\to +\infty$, $a(t_-)\to 0$ and $H(t_-)\to -\beta^{+}$. On the other hand, the Pseudo-Rip in the expanding de Sitter (i.e. EPP), $t_+\to +\infty$, $a(t_+)\to \infty$ and $H(t_+)\to \beta^{+}$, is naturally matched with the contracting de Sitter (i.e. CPP), $t_-\to -\infty$, $a(t_-)\to \infty$ and $H(t_-)\to -\beta^{-}$.

For the moment the whole analysis remains at the kinematic level, namely at the investigation of the required Hubble function evolution. Nevertheless, one should consider the phase portrait \eqref{unified-PhasePortrait} as a field equation governing the dynamics of FLRW cosmology. Therefore, one should provide the theory, namely a modified gravity, that could dynamically produce such a Pseudo-Bang scenario. Amongst the different classes of gravitational modification there is one, i.e. $f(T)$ gravity, which has second-order fields equations and thus it allows for the construction of the phase portrait in a simple way \cite{Awad:2017yod} (in the particular case of Pseudo-Bang scenario the phase portrait 
should be \eqref{unified-PhasePortrait}). Hence, in the following section we investigate the realization of the Pseudo-Bang scenario in the framework of $f(T)$ gravity. Notably, the junction conditions in $f(T)$ gravity has been investigated in \citep{delaCruz-Dombriz:2014zaa}, and used to match two branches of a phase portrait, see \citep{Awad:2017sau}. However, in our case, the matching between the EPP and CPP is naturally fulfilled with no need of junction conditions.

\section{Pseudo-Bang  scenario in $f(T)$ gravity}
\label{Sec:4}

In this section we investigate the  dynamical realization of the Pseudo-Bang
cosmological scenario in the framework of $f(T)$ gravity. We first present a
brief review of the latter and then we proceed to the analysis of its
cosmological implications.

\subsection{$f(T)$ gravity and cosmology}
\label{Sec:4.1}

In the torsional formulation of gravity it proves convenient to  use as
dynamical variables the vierbeins fields ${\mathbf{e}_A(x^\mu)}$, which
at each
manifold point  $x^\mu$ form an orthonormal
basis. In a coordinate basis they can be expressed  as
$\mathbf{e}_A=e^\mu_A\partial_\mu $, related to   the metric through
\begin{equation}  \label{metricrel}
g_{\mu\nu}(x)=\eta_{AB}\, e^A_\mu (x)\, e^B_\nu (x),
\end{equation}
with Greek  and Latin indices respectively running  over coordinate and
tangent space. One introduces the
 Weitzenb\"{o}ck connection
$\overset{\mathbf{w}}{\Gamma}{^\lambda}_{\nu\mu}\equiv e^\lambda_A\:
\partial_\mu
e^A_\nu$ \cite{Weitzenb23}, and therefore the resulting torsion tensor
reads
\begin{equation}
\label{torsten}
{T}^\lambda_{\:\mu\nu}\equiv\overset{\mathbf{w}}{\Gamma}{^\lambda}_{\nu\mu}-%
\overset{\mathbf{w}}{\Gamma}{^\lambda}_{\mu\nu}
=e^\lambda_A\:(\partial_\mu
e^A_\nu-\partial_\nu e^A_\mu).
\end{equation}
This torsion tensor contains all the geometrical information, and thus it
describes the gravitational field. Through its
contraction one acquires the torsion scalar
\begin{equation}
\label{torsiscal}
T\equiv\frac{1}{4}
T^{\rho \mu \nu}
T_{\rho \mu \nu}
+\frac{1}{2}T^{\rho \mu \nu }T_{\nu \mu\rho }
-T_{\rho \mu }^{\ \ \rho }T_{\
\ \ \nu }^{\nu \mu },
\end{equation}
which is then used as  the Lagrangian of the theory.
Variation of the   action in terms of the vierbeins gives identical
equations with general relativity, and that is why this theory was named
  teleparallel
equivalent of general relativity   \cite{Pereira.book}.

Inspired by the curvature modifications of gravity, in which one generalizes
the Einstein-Hilbert action, one can start from teleparallel
equivalent of general relativity  and extend the Lagrangian to an arbitrary
function   of the torsion scalar $T$. The resulting $f(T)$ gravity is
characterized by the action
\cite{Cai:2015emx}
\begin{equation}\label{action}
{\cal S}=\int d^{4}x~ e\left[\frac{1}{2 \kappa^2}f(T)+L_{m}\right],
\end{equation}
with  $e = \text{det}(e_{\mu}^A) = \sqrt{-g}$,
$\kappa^2=8 \pi G$ the gravitational constant, and
where for completeness we have added the matter Lagrangian $L_{m}$. Variation
of the action
(\ref{action}) with respect to the vierbein gives \cite{Cai:2015emx}
\begin{equation}
\label{field_eqns}
\frac{1}{e} \partial_\mu \left( e \gamma_a^{\verb| |\mu\nu} \right)
f_{T}-e_a^\lambda
T^\rho_{\verb| |\mu \lambda}
\gamma_\rho^{\verb| |\nu\mu}
f_{T}+\gamma_a^{\verb| |\mu\nu} \partial_\mu T
f_{TT}
+\frac{1}{4} e_a^\nu f(T)=\frac{{\kappa}^2}{2} e_a^\mu
\mathfrak{T}_\mu^{\verb||\nu},
\end{equation}
with $f_{T}:=\frac{df}{dT}$ and $f_{TT}:=\frac{d^{2}f}{dT^2}$.
Additionally, we have defined
   the superpotential
\begin{equation}\label{superpotential}
\gamma_{\alpha}{^{\mu \nu}}=\frac{1}{4}\left(T_{\alpha}{^{\mu
\nu}}+T{^\mu}{_\alpha}{^\nu}-T{^\nu}{_\alpha}{^\mu}\right)
+\frac{1}{2}\left(\delta^{\nu}_{\alpha}T^{\mu}-\delta^{\mu}_{\alpha}T^{\nu}
\right),
\end{equation}
which is skew symmetric in the last pair of indices, as well as  the
 energy-momentum tensor of the total
matter fields  (baryonic and dark matter and radiation)
${\mathfrak{T}_{\mu}}^{\nu}=e{^a}{_\mu}\left(-\frac{1}{e}\frac{\delta
\mathcal{L}_{m}}{\delta e{^a}{_\nu}}\right)$, assumed  it to be of a perfect
fluid form
\begin{equation}\label{matter}
\mathfrak{T}_{\mu\nu}=\rho u_{\mu}u_{\nu}+p(u_{\mu}u_{\nu}-g_{\mu\nu}),
\end{equation}
with $u_{\mu}=\delta_\mu^t$   the fluid 4-velocity, and $\rho$ and $p$   the
energy density and pressure  in its rest frame.

In order to apply  $f(T)$ gravity to  a cosmological framework  we
consider  the    flat Friedmann-Lema\^itre-Robertson-Walker (FLRW)
geometry
\begin{equation}\label{FRW-metric}
ds^2 = dt^{2}-a(t)^{2}\delta_{ij} dx^{i} dx^{j},
\end{equation}
which corresponds to the vierbein choice
$e_{\mu}^A={\text diag}(1,a,a,a)$,
with $a(t)$ the scale factor (we  impose the natural units   $c=\hbar=k_{B}=1$,
while $\kappa=1/M_{p}$ with
$M_{p}=2.4 \times 10^{18}$ GeV   the reduced Planck mass).
Inserting
into (\ref{field_eqns}) we extract the Friedmann equations
as
\begin{eqnarray}\label{FR1T}
&&H^2= \frac{\kappa^2}{3}\rho
-\frac{f}{6}+\frac{Tf_T}{3},\\ \label{FR2T}
&&\dot{H}=-\frac{\kappa^2(\rho+  p)}{2(1+f_{T}+2Tf_{TT})},
\end{eqnarray}
where we have also made use of the  useful relation that provides the torsion
scalar in FLRW geometry, namely
\begin{equation}\label{Torsionscalar}
    T(t)=-6\left(\frac{\dot{a}}{a}\right)^{2}=-6 H^{2}.
\end{equation}
Additionally, assuming  a barotropic equation of state for the matter fields
of the form (\ref{linear-EoS}), namely $p=(\gamma-1)\rho$, the
matter conservation   equation reads
\begin{equation}\label{continuity}
    \dot{\rho}+3 \gamma H \rho=0.
\end{equation}
Note that the above Friedmann equations for $f(T)=T-2\Lambda$ (with $\Lambda>0$) coincide with
$\Lambda$CDM cosmology.

As it was discussed in the Introduction, one of the advantages of $f(T)$ gravity
is that the Friedmann equation (\ref{FR1T}) gives $\rho\equiv \rho(H)$. Consequently, the differential
equation (\ref{FR2T})  represents a \textit{one-dimensional autonomous
system} of the form $\dot{H}\equiv \mathcal{F}(H)$ as long as the equation of state $p\equiv p(\rho)$ is assumed.
Hence, we can  always interpret it as a vector field on a line, applying one
of the basic techniques of dynamics  by drawing
$\dot{H}$ versus $H$, which helps to analyze the cosmic model in a clear and
transparent way even without solving the system. In order to fix our notation
we follow \cite{book:Steven} calling the above equation   the
\textit{phase portrait}, while its solution $H(t)$ is the \textit{phase
trajectory}. Thus, the phase portrait corresponds to any theory which can be
drawn in an ($\dot{H}-{H}$) \textit{phase-space}. In
this space each point is a \textit{phase point} and could serve as an initial
condition.

Let us now extract the phase portrait of $f(T)$ gravity.   Inserting
(\ref{Torsionscalar}) into (\ref{FR1T}) and (\ref{FR2T}) we can express the
total energy density and pressure  for the matter fields as
\begin{eqnarray}
  \rho &=& \frac{1}{2\kappa^2}\left[f(H)-H f_{H}\right], \label{FR1H}\\
  p &=& \frac{-1}{2\kappa^2}\left[f(H)-H f_{H}-\frac{1}{3}\dot{H}
f_{HH}\right]=\frac{1}{6\kappa^2}\dot{H} f_{HH}-\rho,\qquad\label{FR2H}
\end{eqnarray}
where $f_{H}:=\frac{df}{dH}$ and $f_{HH}:=\frac{d^{2}f}{dH^2}$. Thus, using
additionally the above linear equation of state  we obtain
the
phase portrait equation for any $f(T)$ theory as
\begin{equation}\label{phasetrajectory}
    \dot{H}=3\gamma \left[\frac{f(H)-H
f_{H}}{f_{HH}}\right]\equiv\mathcal{F}(H).
\end{equation}
This shows that the modified Friedmann equations of $f(T)$ gravity
represent a one-dimensional autonomous system
\cite{Awad:2017yod,ElHanafy:2017xsm}, and hence  the theory
in suitable for a phase portrait analysis. It is obvious that the
phase
portrait (\ref{phasetrajectory}) reduces to the $\Lambda$CDM portrait  by setting $f(H)=-6H^2-2\Lambda$.

\subsection{Pseudo-Bang  scenario realization}
\label{Sec:4.4}

In Section \ref{Sec:2} and  Section \ref{Sec:3} we presented the Pseudo-Bang
cosmological scenario, offering a kinematical picture based on the
corresponding
phase space structure. In this subsection we desire to
 reconstruct the $f(H)$ (or equivalently $f(T)$) form which
generates this Pseudo-Bang
  scenario  phase portrait.

One could extract the corresponding phase portrait
  by inserting (\ref{unified-PhasePortrait}) into (\ref{phasetrajectory}).
However, in the Pseudo-Bang  scenario the phase portrait is a double valued
function, thence we expect   two behaviors of $f(H)_{\pm}$ for each Hubble
value, one for the $\dot{H}>0$ branch which we denote   by plus sign, and one
for $\dot{H}<0$ branch which is labeled by negative sign. We mention that at a
particular    Hubble value $H$  the corresponding two values $f(H)_+$ and
$f(H)_-$ characterize two different instants. Thus, in order to avoid the
complexity of the double valued behaviour of the $f(H)$ function at this
moment,
we first evaluate it as a function of cosmic time, i.e. $f(t)$, which in this
case is
monotonic. Inserting the chain rule $f^{\prime} = \dot{f}/\dot{T},~
f^{\prime\prime} = \left(\dot{T} \ddot{f}-\ddot{T} \dot{f}\right)/\dot{T}^{3}$
in Eqs. (\ref{FR1H}) and (\ref{FR2H}), we rewrite the matter density and
pressure as
\begin{eqnarray}
  \rho &=& \frac{1}{2\kappa^2}\left[f(t)-\frac{H}{\dot{H}}\dot{f}(t)\right],
\label{FR1}\\[2pt]
p&=&-\frac{1}{2\kappa^2}\left[f(t)-\left(\frac{H}{\dot{H}}-\frac{\ddot{H}}{3\dot
{H}^{2}}\right)\dot{f}(t)
-\frac{1}{3\dot{H}}\ddot{f}(t)\right].\label{FR2}
\end{eqnarray}
Additionally, the conservation equation (\ref{continuity})
   can be integrated to give
$
\rho=\rho_{0}e^{-3 \int \gamma H dt},
$
where $\rho_{0}\equiv \rho(t_{0})$ is a constant. Recalling Eqs.
(\ref{unified-Hubble}) and (\ref{unified-PhasePortrait}) and  combining with
(\ref{FR1}),  the $f(t)$ function is finally obtained as
\begin{equation}\label{ft0}
f(t)=\left(\beta+\frac{2\alpha \tilde{t}}{2+3\alpha \gamma \tilde{t}^2}\right) \left\{f_0+8\alpha \kappa^2 \rho_0 \int \frac{a_k^{-3\gamma} (\alpha \gamma \tilde{t}^2-2)e^{-3\beta \gamma \tilde{t}}}{(\alpha \gamma \tilde{t}^2+2) \left[(\beta \gamma \tilde{t} + 2)\alpha \tilde{t}+2\beta\right]^2}~d\tilde{t}\right\},
\end{equation}
where $f_{0}$ is an integration constant and $\tilde{t}=t-t_i$.  
One observes that the term associated with the constant $f_0$ is 
$H\propto \sqrt{-T}$, which acts as a divergence term and hence it has no 
contribution to the field equations. Therefore, we omit the $f_0$-term without 
losing the generality of the solution.

In the simple $\beta=0$ case, which corresponds to  the standard bounce
cosmology of (\ref{bounce-scfac}), the above integral can be easily solved and
it produces the $f(T)$ function  obtained  in \cite{Cai:2011tc}. In the general
case of  $\beta\neq 0$, namely in the Pseudo-Bang  cosmology of
(\ref{unified-scfac}), the above integral is found to be
\begin{equation}\label{ft1}
f(t)= \frac{18\kappa^2\rho_{0}
\left[\mathcal{A}e^\theta\Ei(1,\mathcal{B}^+)-\mathcal{A}e^{-\theta}\Ei(1,
\mathcal{B}^-)
+\frac{2}{9}\alpha\theta e^{-3\gamma\beta\tilde{t}}\right]}{\alpha\theta
a_k^{3\gamma}\left(2+3\gamma\alpha\tilde{t}^2\right)},
\end{equation}
where $\theta=\beta\sqrt{-6\gamma/\alpha}$.
In the above expression $\Ei(1,x)$ is the exponential
integral, related to the Gamma function through $\Ei(1,x)=\Gamma(0,x)$,
and
\begin{eqnarray}
\nonumber  \mathcal{A}\equiv \mathcal{A}(\tilde{t}) &=&
\gamma\beta\left(\frac{2}{3}\beta+\frac{2}{3}\alpha \tilde{t}+\alpha\gamma\beta
\tilde{t}^2\right), \\
\nonumber  \mathcal{B}^{\pm}\equiv \mathcal{B}^{\pm}(\tilde{t}) &=&
3\gamma\beta\tilde{t}\pm \theta.
\end{eqnarray}
We mention that the above $f(t)$ function does not exhibit divergences since
$\alpha$ and
$\gamma$ are positive, and hence it is real and finite at all
times  $-\infty < t < +\infty$.

\begin{figure}
\centering
\includegraphics[scale=.52]{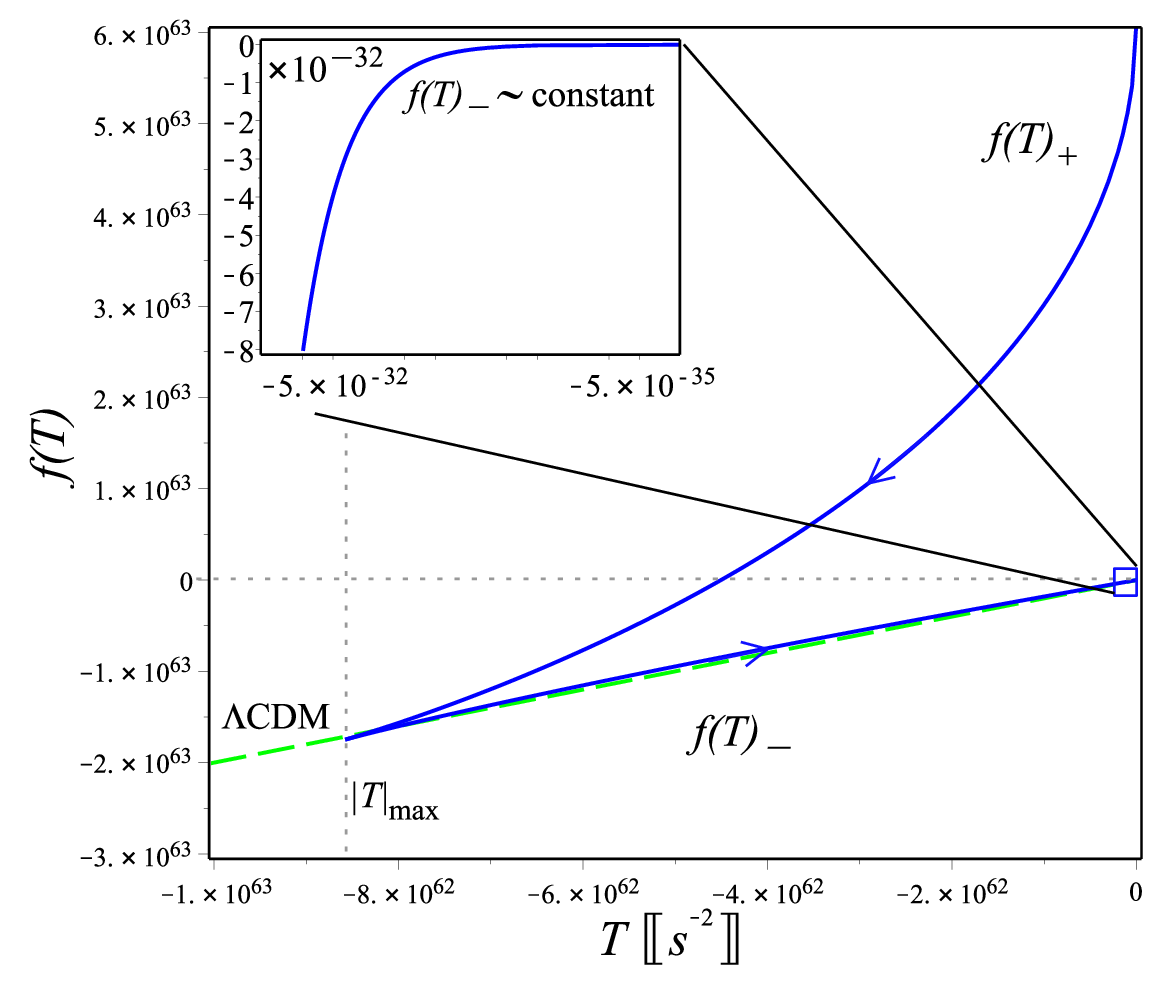}
\caption{{\it{ The reconstructed   $f(T)$ gravity form that
 leads to the realization of the Pseudo-Bang cosmological scenario.
It is a   double-valued function, where $f(T)_+$ gives the evolution
in the positive $\dot{H}$ branch, while $f(T)_-$ gives the evolution in the
negative $\dot{H}$ branch. Note that the  $f(T)_-$ branch practically coincides 
with   $\Lambda$CDM paradigm (green dashed line), except that the latter has no 
$|T|_{\textrm{max}}$. Additionally, $f(T)_\pm$ function is finite at all
times and at late times ($T\sim -6 H_0^2\sim 1.7\times 10^{-35}$
s$^{-2}$) $f(T)_-$   tends to a
constant. The model parameters have been
set as in
(\ref{parameters}).}}}
\label{Fig:f(T)}
\end{figure}

 Using   relation (\ref{unified-time}) which provides the cosmic time as a
function of the Hubble parameter, as well as
(\ref{Torsionscalar}), one can transform $f(t)$ to $f(H)$ and then to
$f(T)$. In particular, we first find
$$\tilde{t}_{\pm}(T)=\frac{\alpha\pm
\sqrt{\alpha^2-6\alpha\gamma\beta^2+2\alpha\beta\gamma\sqrt{-6T}+\alpha\gamma
T}}{3 \gamma \alpha \left(\frac{1}{6}\sqrt{-6T}-\beta\right)},$$
and then
\begin{equation}\label{fT}
    f(T)_\pm=\frac{18\kappa^2\rho_{0}\left(\mathcal{A}(\tilde{t}_\pm)e^\theta\Ei(1,
\mathcal{B}^+(\tilde{t}_\pm))
    -\mathcal{A}(\tilde{t}_\pm)e^{-\theta}\Ei(1,\mathcal{B}^-(\tilde{t}_\pm))
+\frac{2}{9}\alpha\theta e^{-3\gamma\beta\tilde{t}_\pm}\right)}{\alpha\theta
a_k^{3\gamma}\left(2+3\gamma\alpha\tilde{t}_\pm^2\right)},
\end{equation}
which, as mentioned above, is  a double-valued function. This is
one of the  main results of the present work, namely the $f(T)$ form that
generates the
 scale factor (\ref{unified-scfac}), which in turn corresponds to the
Pseudo-Bang  cosmological scenario discussed in Section \ref{Sec:3}.
We mention here that the above form contains the barotropic index of the total
cosmic fluid $\gamma$ as a parameter. This is determined by the dominant
component at each energy scale ($\gamma=4/3$ for radiation epochs while
$\gamma=1$ for matter epoch). For a more realistic and complete expression one
could impose one of the usual unified parameterizations for the total matter
barotropic index that evolves smoothly from $\gamma=4/3$ to $\gamma=1$ during
the cosmological evolution \cite{Chavanis:2012pd}.
For completeness, in Fig. \ref{Fig:f(T)}  we depict this $f(T)$ function
versus  $T$. As we can see it remains  finite, in consistency with the
non-singular universe evolution provided by the Pseudo-Bang  scenario, while
at late times    it practically becomes
constant and the scenario resembles general relativity with a cosmological
constant. Finally, let us point out that our calculations exhibit a continuous behavior,
hence ``small'' deviations from the above $f(T)$ form will only lead to ``small'' deviations from the
resulting scale-factor evolution.

Before closing this Section we make some comments based on  the
equation-of-state behavior. This can be used to investigate the various phantom
crossings and the quintom realization in the present scenario, as well as to
verify its final phase as a Pseudo-Rip one.

Observing the Friedmann equations (\ref{FR1T}), (\ref{FR2T}) we can define an
effective sector of torsional gravitational origin, characterized by an energy
density and pressure of the form
\begin{eqnarray}
    \rho_{T}&=&\frac{1}{2\kappa^2}\left[H
f_{H}-f(H)+6H^{2}\right]=\frac{1}{2\kappa^2}\left[\frac{H}{\dot{H}}
\dot{f}-f(t)+6H^{2}\right],\label{rhoT}\\
    p_{T}&=&-\frac{1}{6\kappa^2}\dot{H}\left(12+f_{HH}\right)-\rho_{T}
=-\frac{1}{6\kappa^2}\left(12\dot{H}-\frac{\ddot{H}}{\dot{H}^2}\dot{f}+\frac{
\ddot{f}}{\dot{H}}\right)-\rho_{T}.\label{pT}
\end{eqnarray}
The matter conservation implies conservation of the above torsional sector too,
namely
 \begin{equation}\label{Tor-continuity}
    \dot{\rho}_{T}+3H\left(1+w _{T}\right)\rho_{T}=0,
\end{equation}
where we have defined the corresponding   equation-of-state parameter
 as
\begin{equation}\label{Tor_equation of state}
w _{T} \equiv \frac{p_{ T}}{\rho_{
T}}=-1+\frac{1}{3}\frac{\dot{H}(\ddot{f}+12\dot{H}^{2})-\ddot{H}\dot{f}}{\dot{H}
\left[\dot{H}\left(f(t)-6H^2\right)-H\dot{f}\right]}.
\end{equation}
Finally, it proves convenient to introduce the total, or effective,
energy density and pressure through
\begin{eqnarray}
&&\rho_{eff}\equiv  \rho+  \rho_{ T}  \label{MFR1}\\
&&p_{eff} \equiv p+p_{ T},\label{MFR2}
\end{eqnarray}
and thus the total
equation-of-state parameter of the universe becomes
\begin{equation}\label{eff_equation of state}
w_{eff}\equiv \frac{p+p_{T}}{\rho+\rho_{T}}=-1-\frac{2}{3}\frac{\dot{H}}{H^2}.
\end{equation}

A first observation is that at the initial Pseudo-Bang  phase we have
$\lim_{t\to
-\infty} w_{T}=\gamma-1$, while at the final phase we obtain $\lim_{t\to
\infty}
w_{T}=-1$. Proceeding forward we will examine the points at which the torsional
equation of state diverges (despite the fact that  for the scenario at hand,
$H$, $\dot{H}$ and $\ddot{H}$ are always finite).
This type of singularity is similar to Type II (sudden)
singularity \cite{Nojiri:2005sx}.
From   (\ref{Tor_equation of state}) we deduce that
$w_{T}\to \pm \infty$ at the points at which $\dot{H}=0$ or
$\frac{\dot{f}}{\dot{H}}=\frac{f}{H}-6H$. According to the first condition, the
torsional fluid is singular at the fixed points $E$ and $G$. We exclude the
special case of the fixed point $A$, since it is not reachable at a finite
time.
Consequently, $w_{T}$ diverges at
$t_{E}=2t_{i}=-2\sqrt{ 2/(3\gamma\alpha)}$ and at $t_{G}=0$. Using
(\ref{unified-Hubble}), (\ref{unified-Hubble-dr}) and (\ref{ft0}), the second
condition above can be solved numerically
to identify one further singular point of $w_{T}$  at $t_{s}\sim
1.2\times 10^{17}$ s, having assumed that  matter is dominated by dust
$\gamma=1$ at this epoch (other choices will not qualitatively alter the
behaviour but just delay the singularity to later times). Hence, at $t_{s}$ the
torsional effective fluid transits from the quintessence ($-1<w_{T}$) to the
phantom regime ($w_{T}<-1$). At a later stage, namely around $t\sim2.2\times
10^{17}$ s, it evolves from phantom into quintessence
regime, and then    evolves asymptotically  towards the cosmological
constant value, namely $w_{T}\to -1$ as $t\to \infty$, which is the Pseudo-Rip
fate. Hence,  in the scenario at hand we have the  quintom realization.
Note that  if $w_{T}$ had remained in the phantom regime and was approaching
from
there the cosmological constant boundary asymptotically, then the universe would
result to a  Little-Rip instead of the Pseudo-Rip phase.

Concerning the total (effective) equation of state  we can easily see from
(\ref{eff_equation of state}) that $
    \lim_{t\to \pm\infty} w_{eff}=-1$.
Additionally, since $\dot{H}$ is always finite it becomes  singular at
the points where $H=0$, namely  at the turnaround point $C$ and the
bounce point $F$. On the other hand, at the de Sitter points $E$ and $G$,
$w_{eff}$ crosses the phantom divide   smoothly.

\begin{figure}
\centering
\includegraphics[width=\textwidth]{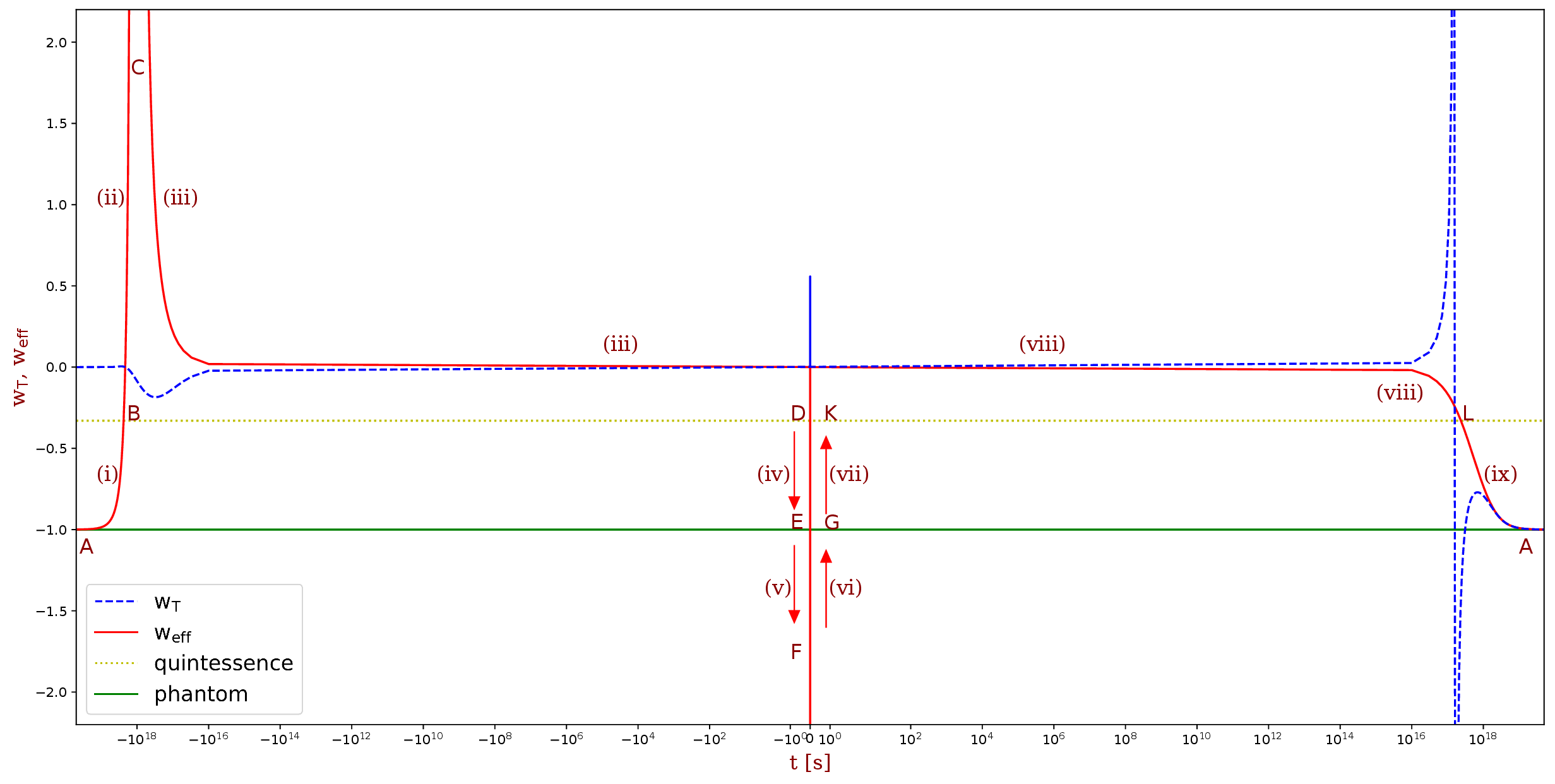}
\caption{{\it{
The evolution of the torsional equation-of-state parameter $w_T$ given in
(\ref{Tor_equation of state}), and of the total, or effective,
equation-of-state parameter $w_{eff}$ given in (\ref{eff_equation of state}),
following  the key points of the Pseudo-Bang scenario.
Additionally,
we draw the quintessence boundary
(at $w=-1/3$) and the phantom divide (at $w=-1$)  in order  to facilitate an
easy comparison with the phase portrait
\ref{Fig:unified-bounce}\subref{fig:unified-phport}.  The model parameters have
been set as in (\ref{parameters}).}}}
\label{Fig:EoS}
\end{figure}

In order to provide a more transparent picture of the   behavior of $w_{T}$
and $w_{eff}$, we depict them in Fig.
\ref{Fig:EoS} for the whole universe evolution, focusing  additionally on the
interval
around 0, and keeping the notation of Fig.
\ref{Fig:unified-bounce}\subref{fig:unified-phport}.
At the Pseudo-Bang  limit, the effective EoS becomes asymptotically a
cosmological
constant, i.e. $w_{eff}\to -1$. In the interval (i) we have  $-1<w_{eff}<-1/3$,
which represents  an inflationary
era. In interval (ii) we have $w_{eff}>-1/3$, and the universe expands with
deceleration. However, the effective fluid $w_{eff}\to +\infty$ at the
turnaround point $C$. In the interval (iii), $w_{eff}\to -1/3$, and the
universe contracts with   deceleration. In the interval (iv), $-1< w_{eff}<
-1/3$, which represents an accelerated contraction era.  Moreover, at point
$E$, the fixed point, we acquire the realization of the phantom-divide   crossing from
non-phantom to phantom regime. In the interval (v), at point $F$, the universe
experiences the bounce, where
$w_{eff}\to -\infty$. After the bounce, in the  interval (vi), the effective
fluid evolves
as $w_{eff}\to -1$ in order to match the observable universe, where the
phantom-divide crossing   occurs at point $G$. In the interval (vii)  the
effective
fluid turns to quintessence regime $-1< w_{eff}<-1/3$, while in the interval
(viii)
the universe gracefully exits into the decelerated   era, where
$-1/3<w_{eff}<w$. In the interval (ix) the effective fluid returns back to the
quintessence regime, where $-1< w_{eff}< -1/3$. Finally, at the Pseudo-Rip
limit  $w_{eff}$ acts asymptotically as a cosmological constant, similarly
to its initial phase.

Finally, let us make some comments on the identification of the last phase in
the universe evolution as Pseudo-Rip, using the energy conditions and the
inertial force interpretation, namely applying the same analysis that  was used
in  \cite{Frampton:2011aa} where the term Pseudo-Rip was first introduced.
At the Pseudo-Bang  limit, $t\to -\infty$, we have
\begin{eqnarray}
 & &\rho\to \infty, p\to \infty,\quad 1 \leq \gamma \leq 2.  \\
  && \rho_{T}\to -\infty, p_{T}\to -\infty,\quad w_{T}\to
\gamma-1. \\
  && \rho_{eff}\to 3\beta^2/\kappa^2,~ p_{eff}\to -3\beta^2/\kappa^2,~ \quad
w_{eff}\to -1.
\end{eqnarray}
Similarly, at the Pseudo-Rip limit, $t\to \infty$, we find
\begin{eqnarray}
 & &\rho\to 0, p\to 0, \quad 1 \leq \gamma \leq 2.  \\
  && \rho_{T}\to 3\beta^2/\kappa^2, p_{T}\to -3\beta^2/\kappa^2,\quad w_{T}\to
-1. \\
  && \rho_{eff}\to 3\beta^2/\kappa^2,~ p_{eff}\to -3\beta^2/\kappa^2,~ \quad
w_{eff}\to -1.
\end{eqnarray}
Hence, the total energy density at asymptotically early and late times is
non-zero, which is consistent with the fact that $H$ is constant at these
phases. Although the energy scales at Pseudo-Bang and at Pseudo-Rip are 
almost equivalent, the scale factor and matter sector take two different 
extremes. As $H\to \beta^{-}$, at Pseudo-Bang, the scale factor $a\to 0$, the 
matter density ($\rho \propto a^{-3}$), i.e. $\rho\to \infty$, the matter 
pressure $p\to \infty$ and the temperature $\Theta\to \infty$ for $\gamma>1$. As 
$H\to \beta^{+}$, at Pseudo-Rip, the scale factor $a\to \infty$, the matter 
density $\rho \to 0$, the matter pressure $p\to 0$ and the temperature 
$\Theta\to 0$.  Hence, these two phases are not similar, as one 
might think by comparing only their energy scales.
\begin{figure}[ht]
\centering
\subfigure[~Full space
evolution]{\label{fig:force1}\includegraphics[scale=.43]{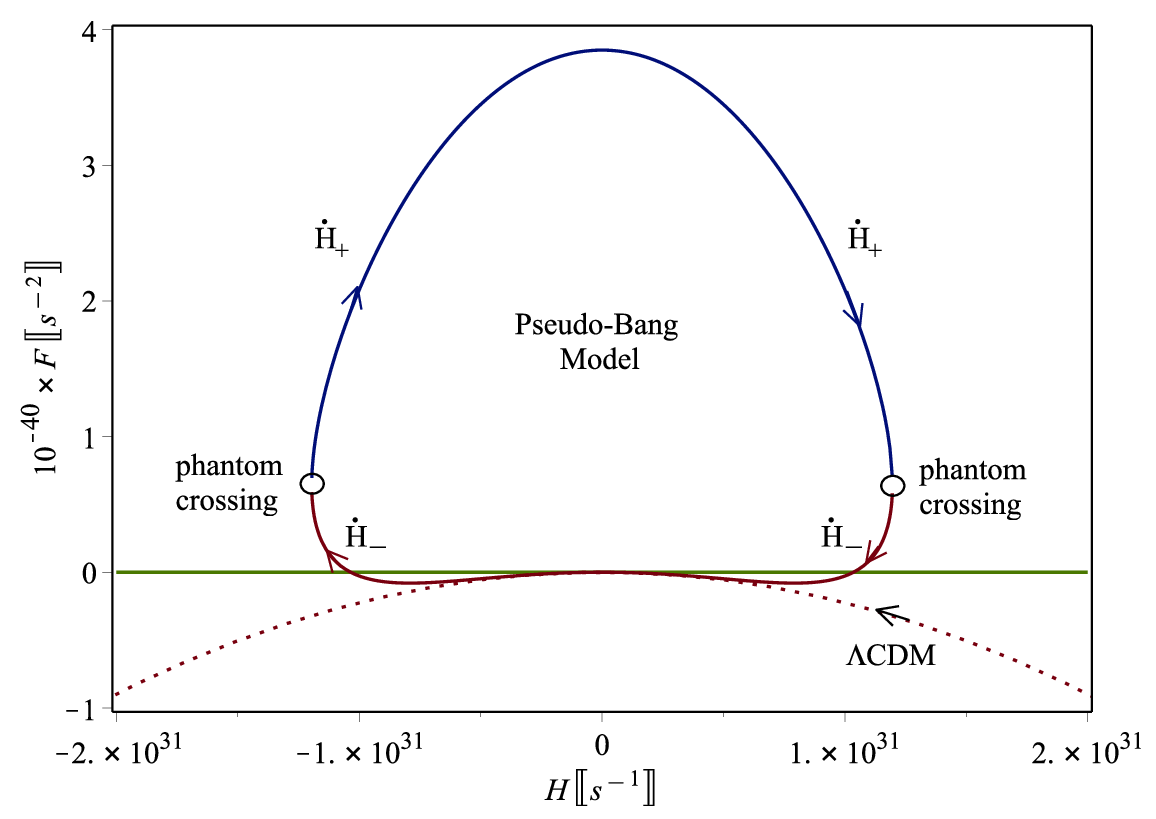}}\hspace{
0.1cm}
\subfigure[~Future
ripping]{\label{fig:force2}\includegraphics[scale=.35]{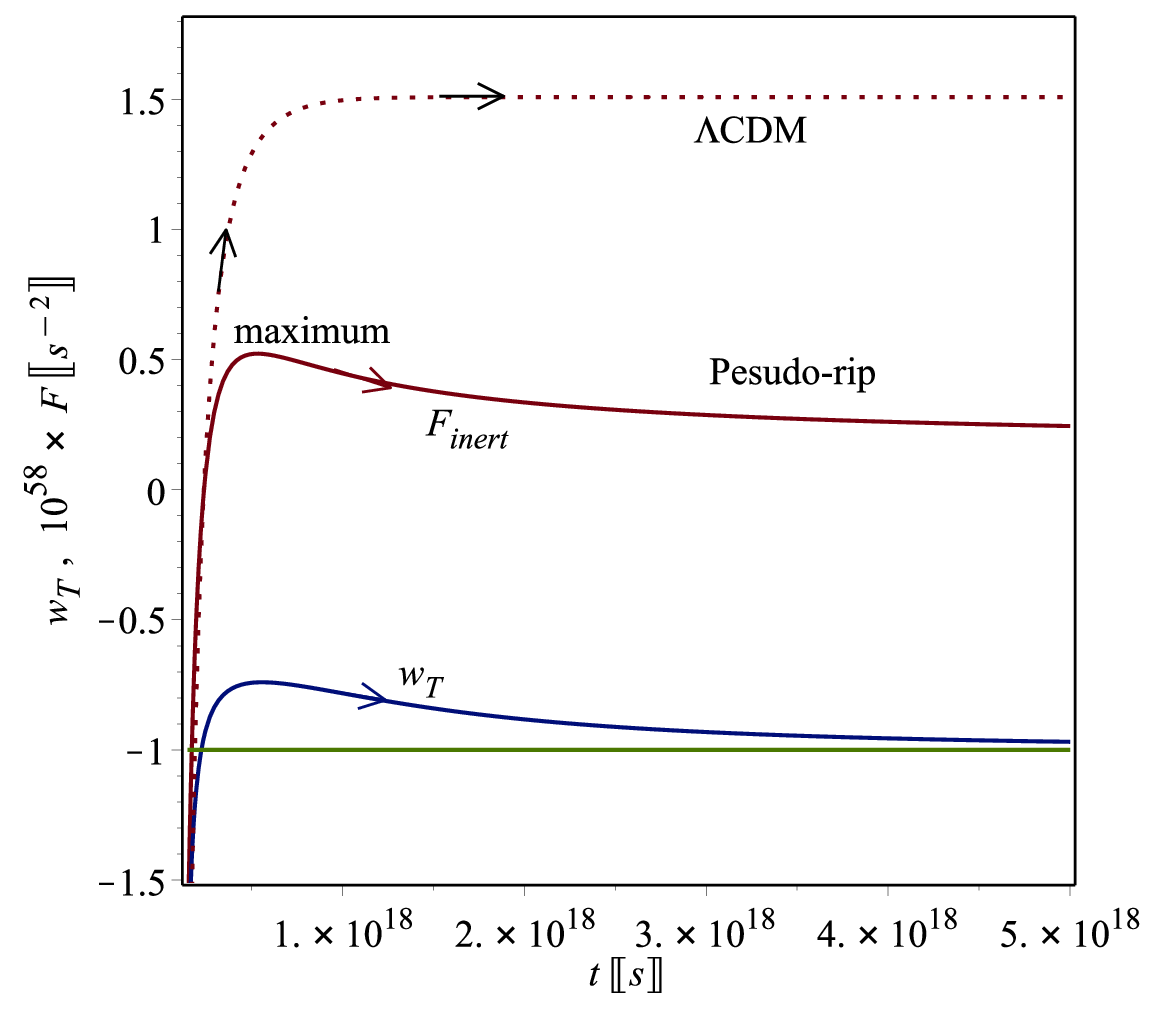}}
\caption[figtopcap]{
\subref{fig:force1} {\it The ripping inertial force (\ref{force}) versus the
Hubble function for the whole Pseudo-Bang scenario and for $\Lambda$CDM. It
is finite at all phases, and the crossing of the phantom divide is achieved through fixed points that can be reached in a finite time,
characterized by $dF_{inert}/dH\to \infty$.}
 \subref{fig:force2}
 {\it The ripping inertial force as a function of time in the current and future
universe for Pseudo-Bang and $\Lambda$CDM scenarios. The inertial force in the Pseudo-Bang exhibits a maximum and then decreases asymptotically to a constant value as required by the Pseudo-Rip phase which may allow some structures to rebuilt, while in $\Lambda$CDM the force increases monotonically to a constant value. The model parameters have been set as in (\ref{parameters}).}}
\label{Fig:inertial-force}
\end{figure}

Concerning  the inertial (ripping) force, for any two points separated by a
comoving distance $l$, the relative acceleration between them is $l\ddot{a}/a$.
Thus, the inertial force on a mass $m$ as seen by an observer at a comoving
distance $l$ is given by \cite{Frampton:2011rh,Frampton:2011aa}
\begin{equation}\label{force}
    F_{inert}=ml\frac{\ddot{a}}{a}=ml(\dot{H}+H^2).
\end{equation}
Using the phase portrait (\ref{unified-PhasePortrait}), in Fig.
\ref{Fig:inertial-force}\subref{fig:force1}  we draw the
  inertial force as a function of the Hubble parameter. As we observe,   the
inertial force is very sensitive to  the  Hubble parameter changes at
the phantom crossing points, since $dF_{inert}/dH\to \infty$ at these fixed points.
Furthermore, in Fig. \ref{Fig:inertial-force}\subref{fig:force2} we depict  the
inertial force time evolution, from which we can see that it will peak at a
future time $\sim 5.3 \times 10^{17}$ s and then it will decrease
asymptotically to a fixed limit $F_{inert}\to ml\beta^2$. Such a behavior, is
the definition of a Pseudo-Rip fate  \cite{Frampton:2011aa}, which
distinguishes it from   other Rips (in fact it is  an
intermediate between the no Rip and the Little Rip) and shows that  the bound
structures dissociate if they are at or below a particular threshold.
Note that this behavior of $F_{inert}$ is closely related to the behavior of
$w_T$, that is why we have added it in the same Figure for completeness.

\section{Summary and Remarks}\label{Sec:5}

In the present work we   studied   the complete universe
evolution in the framework of $f(T)$ cosmology. As a first step  we
investigated the necessary requirements at the kinematic level in order to
describe  the standard observed thermal history of the universe, namely the
sequence of radiation, matter and late-time acceleration epochs, as well as
being able to bypass the initial singularity. In particular, we
introduced an exponential correction to the standard bouncing scale
factor.

 In order to investigate the   cosmological behavior of the introduced
scale factor, we performed
a detailed analysis of the ($H$, $\dot{H}$)   phase portrait. Firstly, we showed
that    the Minkowskian origin of the standard bounce universe is instead
shifted to a de Setterian origin  in the present
scenario. This allows the universe to begin with an
accelerated expansion phase, since  in the infinite past $t\to -\infty$ the
scale factor goes to 0,  the Hubble parameter goes to a constant, and its
derivative to  $0^{-}$. Since these features resemble those of the Pseudo-Rip
fate \cite{Frampton:2011aa}  but in a revered way, we called the initial phase
as    Pseudo-Bang and the whole cosmological scenario as Pseudo-Bang scenario.

After the  Pseudo-Bang, the universe evolves in a first inflationary phase, a
cosmological turnaround and then experiences a bounce, after which we have a
second inflationary regime with a successful exit. Then the universe follows
the standard thermal history of the sequence of radiation, matter and late-time
acceleration epochs. Finally, the   last phase as $t\to \infty$
is an everlasting Pseudo-Rip one, where
the  scale factor goes to infinity,  the Hubble parameter goes to a constant,
and its derivative to  $0^{-}$. Interestingly enough,
 the origin and the fate of
the universe are characterized by the same energy scale, $E\sim \sqrt{M_p
H}\sim   10^{-12}$ GeV. Additionally, after the turnaround and after the bounce
the universe exhibits  the crossing
of the phantom divide twice through a particular type of fixed points at which the slope $d\dot{H}/dH$ becomes infinite making them reachable in a finite time. These have almost
the same energy scale, $E \sim 4\times 10^{12}$ GeV, with a time separation
$\sim 10^{-32}$ s, in a phantom regime. It is this second phantom crossing that
 provides the second inflationary realization.
We  examined the evolution of the primordial fluctuations
versus the Hubble radius, and we showed that all fluctuations are initially
sub-horizon, which is consistent with the observations of acoustic peaks in the
power spectrum of CMB anisotropies. This allows to naturally assume
Bunch-Davies vacuum conditions of the quantum fluctuations around the initial
vacuum state. Interestingly, the Pseudo-Bang phase resembles de Sitter case up to a sub-leading term, whereas the hypersurface is not space-like but null. In this sense, the Pseudo-Bang phase is not a genuine beginning and it can be preceded by a contraction phase at $H=-\beta$.

Having investigated the kinematic requirements for the Pseudo-Bang
cosmological scenario we examined its    dynamical realization in the framework
of $f(T)$ gravity. We  took advantage of the fact that the field equations of
$f(T)$ gravity are of second order, and therefore the corresponding autonomous
dynamical  system is  one dimensional. This results to a phase space in the
($H$, $\dot{H}$) plane that can incorporate the aforementioned kinematic features
of the  Pseudo-Bang scale factor. Hence,  we reconstructed the specific $f(T)$
form that can lead to the  Pseudo-Bang cosmological scenario.  Lastly, by
studying the torsional and the total equation-of-state  parameters we saw that
the total, effective fluid does not exhibit any singular behaviour at the
phantom  crossing points, while the   torsional fluid feels
them as Type II singular phases.

In conclusion, we provided an $f(T)$ form that can generate the complete
universe evolution  from  Pseudo-Bang to  Pseudo-Rip, including the
standard thermal history. It would be interesting to perform a full
observational confrontation using data from  Supernovae type Ia
   (SNIa), Baryonic Acoustic Oscillations (BAO), Cosmic Microwave Background
(CMB) shift parameter, and Hubble parameter measurements, in order to examine
whether the Pseudo-Bang scenario at late times is in agreement with
observations. Moreover, we should perform a detailed perturbation analysis 
(allowing also for possoble ``soft'' features \cite{Saridakis:2021qxb})  in
order to confront it with CMB temperature and polarization data, as well as
with $f\sigma_8$ and growth data. Such investigations, although necessary,
lie beyond the scope of the present work and are left for   future
projects.

\begin{acknowledgments}
We would like to thank the referees for valuable suggestions which improved the 
paper. W.E. gratefully acknowledges the technical assistance of M. Hashim to 
create the graphs of Fig. \ref{Fig:EoS}, and wishes to thank Adel Awad for 
helpful discussions about 
Penrose-Carter diagram. E.N.S work is supported in part by the USTC Fellowship 
for international professors.

\end{acknowledgments}





\begin{thebibliography}{100}






\bibitem{Bartolo:2004if}
  N.~Bartolo, E.~Komatsu, S.~Matarrese and A.~Riotto,
    {\it{Non-Gaussianity from inflation: Theory and observations}},
  Phys.\ Rept.\  {\bf 402}, 103 (2004)
  [\href{http://xxx.lanl.gov/abs/astro-ph/0406398}
{{\tt arXiv:astro-ph/0406398}}].



\bibitem{Olive:1989nu}
  K.~A.~Olive,
     {\it{Inflation}},
  Phys.\ Rept.\  {\bf 190}, 307 (1990).




\bibitem{Copeland:2006wr}
  E.~J.~Copeland, M.~Sami and S.~Tsujikawa,
  {\it{Dynamics of dark energy}},
  Int.\ J.\ Mod.\ Phys.\  D {\bf 15}, 1753 (2006)
     [\href{http://xxx.lanl.gov/abs/hep-th/0603057}
{{\tt arXiv:hep-th/0603057}}].



\bibitem{Cai:2009zp}
  Y.~-F.~Cai, E.~N.~Saridakis, M.~R.~Setare and J.~-Q.~Xia,
  {\it{Quintom Cosmology: Theoretical implications and observations}},
  Phys.\ Rept.\  {\bf 493}, 1 (2010)
      [\href{http://xxx.lanl.gov/abs/0909.2776}
{{\tt arXiv:0909.2776}}].

 



\bibitem{Capozziello:2011et}
  S.~Capozziello and M.~De Laurentis,
 {\it{Extended Theories of Gravity}},
  Phys.\ Rept.\  {\bf 509}, 167 (2011)
[\href{http://xxx.lanl.gov/abs/1108.6266}
{{\tt arXiv:1108.6266}}].



\bibitem{CANTATA:2021ktz}
E.~N.~Saridakis \textit{et al.} [CANTATA],
[\href{http://xxx.lanl.gov/abs/2105.12582}
{{\tt arXiv:2105.12582}}].

 





\bibitem{DeFelice:2010aj}
  A.~De Felice and S.~Tsujikawa,
  {\it{f(R) theories}},
  Living Rev.\ Rel.\  {\bf 13}, 3 (2010)
  [\href{http://xxx.lanl.gov/abs/1002.4928}
  {{\tt arXiv:1002.4928}}].





\bibitem{Nojiri:2010wj}
  S.~Nojiri and S.~D.~Odintsov,
  {\it{Unified cosmic history in modified gravity: from F(R) theory to
Lorentz
non-invariant
models}},
  Phys.\ Rept.\  {\bf 505}, 59 (2011)
[\href{http://xxx.lanl.gov/abs/1011.0544}
 {{\tt arXiv:1011.0544}}].



\bibitem{Nojiri:2005jg}
  S.~Nojiri and S.~D.~Odintsov,
{\it{Modified Gauss-Bonnet theory as gravitational alternative for dark
energy}},
  Phys.\ Lett.\ B {\bf 631}, 1 (2005)
[\href{http://xxx.lanl.gov/abs/hep-th/0508049}
{{\tt arXiv:hep-th/0508049}}].


\bibitem{Lovelock:1971yv}
  D.~Lovelock,
{\it{The Einstein tensor and its generalizations}},
  J.\ Math.\ Phys.\  {\bf 12}, 498 (1971).




\bibitem{Horndeski:1974wa}
G.~W.~Horndeski,
{\it{Second-order scalar-tensor field equations in a four-dimensional space}},
Int. J. Theor. Phys. \textbf{10}, 363-384 (1974).

\bibitem{Nicolis:2008in}
A.~Nicolis, R.~Rattazzi and E.~Trincherini,
{\it{The Galileon as a local modification of gravity}},
Phys. Rev. D \textbf{79}, 064036 (2009).
[\href{http://xxx.lanl.gov/abs/0811.2197}
{{\tt arXiv:0811.2197}}].





\bibitem{Deffayet:2009wt}
C.~Deffayet, G.~Esposito-Farese and A.~Vikman,
{\it{Covariant Galileon}},
Phys. Rev. D \textbf{79}, 084003 (2009)
[\href{http://xxx.lanl.gov/abs/0901.1314}
{{\tt arXiv:0901.1314}}].





\bibitem{Brax:2004xh}
P.~Brax, C.~van de Bruck and A.~C.~Davis,
{\it{Brane world cosmology}},
Rept. Prog. Phys. \textbf{67}, 2183-2232 (2004)
[\href{http://xxx.lanl.gov/abs/hep-th/0404011}
{{\tt arXiv:hep-th/0404011}}].








 \bibitem{ein28}
A. Einstein,
{\it{Riemannian geometry, while maintaining the notion of teleparallelism}},
  Sitz. Preuss. Akad. Wiss. {\bf 17},  217  (1928).

  \bibitem{ein28a}
A. Einstein,
{\it{A new possibility for a unified field theory of gravitation and
electromagnetism}},
  Sitz. Preuss. Akad. Wiss. {\bf 17},  224  (1928).


  \bibitem{ein28b}
  A.~Unzicker and T.~Case,
  \emph{Translation of Einstein's attempt of a unified field theory with
 teleparallelism},
\href{https://arxiv.org/pdf/physics/0503046.pdf}{[{\tt
{physics/0503046}}]}.



\bibitem{Hayashi79}
  K. Hayashi and T. Shirafuji,
 \emph{New general relativity},
   {Phys. Rev. D\textbf{19}}, (1979)  3524.



\bibitem{Pereira.book}
R.~{Aldrovandi} and J.~G. {Pereira}, \emph{{Teleparallel Gravity}},
\newblock Springer Science \& Business Media Dordrecht (2013).





\bibitem{Maluf:2013gaa}
  J.~W.~Maluf,
{\it{The teleparallel equivalent of general relativity}},
{{Annalen
Phys.\  {\bfseries 525}}, (2013) 339},
[\href{http://xxx.lanl.gov/abs/1303.3897}
{{\tt arXiv:1303.3897}}].



\bibitem{Bengochea:2008gz}
G.~R. Bengochea and R.~Ferraro, \emph{{Dark torsion as the cosmic speed-up}},
  {Phys. Rev. D}
  {\bfseries 79} (2009) 124019,
  [\href{http://xxx.lanl.gov/abs/0812.1205}
{{\tt arXiv:0812.1205}}].




\bibitem{Linder:2010py}
E.~V. Linder,
\emph{Einstein's Other Gravity and the Acceleration of the
  Universe},
  Phys. Rev. D {\bfseries 81} (2010)
  127301,
  [\href{http://xxx.lanl.gov/abs/1005.3039}
{{\tt arXiv:1005.3039}}].










\bibitem{Cai:2015emx}
  Y.~F.~Cai, S.~Capozziello, M.~De Laurentis and E.~N.~Saridakis,
 {\it{f(T) teleparallel gravity and cosmology}},
  Rept.\ Prog.\ Phys.\  {\bf 79}, 106901 (2016)
  [\href{http://xxx.lanl.gov/abs/1511.07586}
{{\tt arXiv:1511.07586}}].





\bibitem{Kofinas:2014owa}
  G.~Kofinas and E.~N.~Saridakis,
  {\it{Teleparallel equivalent of Gauss-Bonnet gravity and its
modifications}},
  Phys.\ Rev.\ D {\bf 90}, 084044 (2014)
  [\href{http://xxx.lanl.gov/abs/1404.2249}
  {{\tt arXiv:1404.2249}}].


\bibitem{Kofinas:2014daa}
G.~Kofinas and E.~N.~Saridakis,
{\it{Cosmological applications of $F(T,T_G)$ gravity}},
Phys. Rev. D \textbf{90}, 084045 (2014)
  [\href{http://xxx.lanl.gov/abs/1408.0107}
{{\tt arXiv:1408.0107}}].





\bibitem{Geng:2011aj}
C.-Q. Geng, C.-C. Lee, E.~N. Saridakis and Y.-P. Wu,
\emph{{Teleparallel dark
  energy}},
 {{Phys.
  Lett. B} {\bfseries 704} (2011) 384--387},
    [\href{http://xxx.lanl.gov/abs/1109.1092}
{{\tt arXiv:1109.1092}}].



\bibitem{Hohmann:2018rwf}
M.~Hohmann, L.~J\"arv and U.~Ualikhanova,
{\it{Covariant formulation of scalar-torsion gravity}},
Phys. Rev. D \textbf{97}, no.10, 104011 (2018)
    [\href{http://xxx.lanl.gov/abs/1801.05786}
{{\tt arXiv:1801.05786}}].







\bibitem{Chen:2010va}
  S.~H.~Chen, J.~B.~Dent, S.~Dutta and E.~N.~Saridakis,
{\it{Cosmological perturbations in f(T) gravity}},
  Phys.\ Rev.\  D {\bf 83}, 023508 (2011)
 [\href{http://xxx.lanl.gov/abs/1008.1250}
{{\tt arXiv:1008.1250}}].







\bibitem{Zheng:2010am}
  R.~Zheng and Q.~G.~Huang,
   {\it{Growth factor in f(T) gravity}},
{{JCAP {\bfseries 1103}, (2011) 002,}}
   [\href{http://xxx.lanl.gov/abs/1010.3512}
{{\tt arXiv:1010.3512}}].



\bibitem{Bamba:2010wb}
K.~Bamba, C.~Q.~Geng, C.~C.~Lee and L.~W.~Luo,
{\it{Equation of state for dark energy in $f(T)$ gravity}},
JCAP \textbf{01}, 021 (2011)
   [\href{http://xxx.lanl.gov/abs/1011.0508}
{{\tt arXiv:1011.0508}}].






\bibitem{Li:2011rn}
  M.~Li, R.~X.~Miao and Y.~G.~Miao,
  {\it{Degrees of freedom of $f(T)$ gravity}},
{{JHEP
{\bfseries
1107}, (2011)  108}},
   [\href{http://xxx.lanl.gov/abs/1105.5934}
{{\tt arXiv:1105.5934}}].






\bibitem{Capozziello:2011hj}
  S.~Capozziello, V.~F.~Cardone, H.~Farajollahi and A.~Ravanpak,
   {\it{Cosmography in f(T)-gravity}},
{{
  Phys.\ Rev.\  D {\bfseries 84}, (2011) 043527}},
     [\href{http://xxx.lanl.gov/abs/1108.2789}
{{\tt arXiv:1108.2789}}].





\bibitem{Wu:2011kh}
  Y.~P.~Wu and C.~Q.~Geng,
     {\it{Primordial Fluctuations within Teleparallelism}},
{{Phys.\ Rev.\ D {\bfseries 86}, (2012)
104058}},
     [\href{http://xxx.lanl.gov/abs/1110.3099}
{{\tt arXiv:1110.3099}}].





\bibitem{Wei:2011aa}
  H.~Wei, X.~J.~Guo and L.~F.~Wang,
     {\it{Noether Symmetry in $f(T)$ Theory}},
{{Phys.\ Lett.\  B {\bfseries 707} , (2012) 298}},
      [\href{http://xxx.lanl.gov/abs/1112.2270}
{{\tt arXiv:1112.2270}}].







 \bibitem{Amoros:2013nxa}
  J.~Amoros, J.~de Haro and S.~D.~Odintsov,
    {\it{Bouncing Loop Quantum Cosmology from $F(T)$
 gravity}},
{{Phys.\ Rev.\ D {\bfseries 87},  (2013) 104037}}
      [\href{http://xxx.lanl.gov/abs/1305.2344}
{{\tt arXiv:1305.2344}}].




 \bibitem{Otalora:2013dsa}
  G.~Otalora,
     {\it{Cosmological dynamics of tachyonic teleparallel dark energy}},
  {{Phys.\
Rev.\ D {\bf
88}, (2013) 063505}},
      [\href{http://xxx.lanl.gov/abs/1305.5896}
{{\tt arXiv:1305.5896}}].






\bibitem{Bamba:2013jqa}
K.~Bamba, S.~D.~Odintsov and D.~S\'aez-G\'omez,
{\it{Conformal symmetry and accelerating cosmology in teleparallel gravity}},
Phys. Rev. D \textbf{88}, 084042 (2013)
      [\href{http://xxx.lanl.gov/abs/1308.5789}
{{\tt arXiv:1308.5789}}].





\bibitem{Li:2013xea}
J.~T.~Li, C.~C.~Lee and C.~Q.~Geng,
{\it{Einstein Static Universe in Exponential $f(T)$ Gravity}},
Eur. Phys. J. C \textbf{73}, no.2, 2315 (2013)
      [\href{http://xxx.lanl.gov/abs/1302.2688}
{{\tt arXiv:1302.2688}}].






\bibitem{Ong:2013qja}
  Y.~C.~Ong, K.~Izumi, J.~M.~Nester and P.~Chen,
   {\it{Problems with Propagation and Time Evolution in f(T) Gravity}},
{{Phys.\
Rev.\ D {\bfseries 88} (2013) 2,  024019}},
      [\href{http://xxx.lanl.gov/abs/1303.0993}
{{\tt arXiv:1303.0993}}].

\bibitem{Paliathanasis:2014iva}
A.~Paliathanasis, S.~Basilakos, E.~N.~Saridakis, S.~Capozziello, K.~Atazadeh,
F.~Darabi and M.~Tsamparlis,
 {\it{New Schwarzschild-like solutions in f(T) gravity through Noether
symmetries}},
Phys. Rev. D \textbf{89}, 104042 (2014)
      [\href{http://xxx.lanl.gov/abs/1402.5935}
{{\tt arXiv:1402.5935}}].




\bibitem{Nashed:2014lva}
G.~G.~L.~Nashed and W.~El Hanafy,
{\it{A Built-in Inflation in the $f(T)$-Cosmology}},
Eur. Phys. J. C \textbf{74}, 3099 (2014)
      [\href{http://xxx.lanl.gov/abs/1403.0913}
{{\tt arXiv:1403.0913}}].





\bibitem{Darabi:2014dla}
F.~Darabi, M.~Mousavi and K.~Atazadeh,
\emph{{Geodesic deviation equation in
  f(T) gravity}},
 {{Phys.
  Rev. D} {\bfseries 91} (2015) 084023},
        [\href{http://xxx.lanl.gov/abs/1501.00103}
{{\tt arXiv:1501.00103}}].






\bibitem{Malekjani:2016mtm}
M.~Malekjani, N.~Haidari and S.~Basilakos,
{\it{Spherical collapse model and cluster number counts in power law $f(T)$
gravity}},
Mon. Not. Roy. Astron. Soc. \textbf{466}, no.3, 3488-3496 (2017)
        [\href{http://xxx.lanl.gov/abs/1609.01964}
{{\tt arXiv:1609.01964}}].







\bibitem{Farrugia:2016qqe}
  G.~Farrugia and J.~L.~Said,
  \emph{Stability of the flat FLRW metric in $f(T)$ gravity},
{{
Phys.\
Rev.\ D {\bfseries 94}, no. 12,  (2016)
124054},}
        [\href{http://xxx.lanl.gov/abs/1701.00134}
{{\tt arXiv:1701.00134}}].






\bibitem{Bahamonde:2017wwk}
  S.~Bahamonde, C.~G.~B\"{o}hmer and M.~ M.~Kr\v{s}\v{s}\'ak,
 \emph{New classes of modified teleparallel gravity models,}
{{Phys.\ Lett.\ B {\bfseries
775},
(2017) 37}
 },
         [\href{http://xxx.lanl.gov/abs/1706.04920}
{{\tt arXiv:1706.04920}}].



\bibitem{Qi:2017xzl}
J.~Z.~Qi, S.~Cao, M.~Biesiada, X.~Zheng and H.~Zhu,
{\it{New observational constraints on $f(T)$ cosmology from radio quasars}},
Eur. Phys. J. C \textbf{77}, no.8, 502 (2017)
         [\href{http://xxx.lanl.gov/abs/1708.08603}
{{\tt arXiv:1708.08603}}].




\bibitem{Karpathopoulos:2017arc}
L.~Karpathopoulos, S.~Basilakos, G.~Leon, A.~Paliathanasis and M.~Tsamparlis,
{\it{Cartan symmetries and global dynamical systems analysis in a higher-order
modified teleparallel theory}},
Gen. Rel. Grav. \textbf{50}, no.7, 79 (2018)
         [\href{http://xxx.lanl.gov/abs/1709.02197}
{{\tt arXiv:1709.02197}}].



\bibitem{Abedi:2018lkr}
H.~Abedi, S.~Capozziello, R.~D'Agostino and O.~Luongo,
{\it{Effective gravitational coupling in modified teleparallel theories}},
Phys. Rev. D \textbf{97}, no.8, 084008 (2018)
         [\href{http://xxx.lanl.gov/abs/1803.07171}
{{\tt arXiv:1803.07171}}].





\bibitem{Krssak:2018ywd}
M.~Krssak, R.~J.~van den Hoogen, J.~G.~Pereira, C.~G.~B\"ohmer and A.~A.~Coley,
{\it{Teleparallel theories of gravity: illuminating a fully invariant
approach}},
Class. Quant. Grav. \textbf{36}, no.18, 183001 (2019)
         [\href{http://xxx.lanl.gov/abs/1810.12932}
{{\tt arXiv:1810.12932}}].




\bibitem{Iosifidis:2018zwo}
D.~Iosifidis and T.~Koivisto,
{\it{Scale transformations in metric-affine geometry}},
         [\href{http://xxx.lanl.gov/abs/1810.12276}
{{\tt arXiv:1810.12276}}].





\bibitem{El-Zant:2018bsc}
A.~El-Zant, W.~El Hanafy and S.~Elgammal,
{\it{$H_0$ Tension and the Phantom Regime: A Case Study in Terms of an Infrared
$f(T)$ Gravity}},
Astrophys. J. \textbf{871} (2019) no.2, 210
        [\href{http://xxx.lanl.gov/abs/1809.09390}
{{\tt arXiv:1809.09390}}].


\bibitem{Anagnostopoulos:2019miu}
F.~K.~Anagnostopoulos, S.~Basilakos and E.~N.~Saridakis,
{\it{Bayesian analysis of $f(T)$ gravity using $f\gamma_8$ data}},
Phys. Rev. D \textbf{100}, no.8, 083517 (2019)
        [\href{http://xxx.lanl.gov/abs/1907.07533}
{{\tt arXiv:1907.07533}}].



\bibitem{Nunes:2019bjq}
R.~C.~Nunes, M.~E.~S.~Alves and J.~C.~N.~de Araujo,
{\it{Forecast constraints on $f(T)$ gravity with gravitational waves from
compact
binary coalescences}},
Phys. Rev. D \textbf{100}, no.6, 064012 (2019)
         [\href{http://xxx.lanl.gov/abs/1905.03237}
{{\tt arXiv:1905.03237}}].




\bibitem{Yan:2019gbw}
S.~F.~Yan, P.~Zhang, J.~W.~Chen, X.~Z.~Zhang, Y.~F.~Cai and E.~N.~Saridakis,
{\it{Interpreting cosmological tensions from the effective field theory of
torsional gravity}},
Phys. Rev. D \textbf{101}, no.12, 121301 (2020)
         [\href{http://xxx.lanl.gov/abs/1909.06388}
{{\tt arXiv:1909.06388}}].

\bibitem{ElHanafy:2019zhr}
W.~El Hanafy and G.~G.~L.~Nashed,
{\it{Phenomenological Reconstruction of $f(T)$ Teleparallel Gravity}},
Phys. Rev. D \textbf{100} (2019) no.8, 083535
        [\href{http://xxx.lanl.gov/abs/1910.04160}
{{\tt arXiv:1910.04160}}].


\bibitem{Saridakis:2019qwt}
E.~N.~Saridakis, S.~Myrzakul, K.~Myrzakulov and K.~Yerzhanov,
{\it{Cosmological applications of $F(R,T)$ gravity with dynamical curvature and
torsion}},
Phys. Rev. D \textbf{102}, no.2, 023525 (2020)
        [\href{http://xxx.lanl.gov/abs/1912.03882}
{{\tt arXiv:1912.03882}}].



\bibitem{Wang:2020zfv}
D.~Wang and D.~Mota,
{\it{Can $f(T)$ gravity resolve the $H_0$ tension?}},
Phys. Rev. D \textbf{102}, no.6, 063530 (2020)
         [\href{http://xxx.lanl.gov/abs/2003.10095}
{{\tt arXiv:2003.10095}}].





\bibitem{Bahamonde:2020lsm}
S.~Bahamonde, V.~Gakis, S.~Kiorpelidi, T.~Koivisto, J.~Levi Said and
E.~N.~Saridakis,
{\it{Cosmological perturbations in modified teleparallel gravity models:
Boundary
term extension}},
         [\href{http://xxx.lanl.gov/abs/2009.02168}
{{\tt arXiv:2009.02168}}].






\bibitem{Briffa:2020qli}
R.~Briffa, S.~Capozziello, J.~Levi Said, J.~Mifsud and E.~N.~Saridakis,
{\it{Constraining Teleparallel Gravity through Gaussian Processes}},
         [\href{http://xxx.lanl.gov/abs/2009.14582}
{{\tt arXiv:2009.14582}}].

\bibitem{Hashim:2020sez}
M.~Hashim, W.~El Hanafy, A.~Golovnev and A.~El-Zant,
{\it{Toward a concordance teleparallel Cosmology I: Background Dynamics}},
        [\href{http://xxx.lanl.gov/abs/2010.14964}
{{\tt arXiv:2010.14964}}].


\bibitem{Hashim:2021pkq}
M.~Hashim, A.~A.~El-Zant, W.~El Hanafy and A.~Golovnev,
{\it{Toward a concordance teleparallel cosmology. Part~II. Linear 
perturbation}},
JCAP \textbf{07} (2021), 053
        [\href{http://xxx.lanl.gov/abs/2104.08311}
{{\tt arXiv:2104.08311}}].


 
  

\bibitem{Borde:1993xh}
  A.~Borde and A.~Vilenkin,
 {\it{Eternal Inflation And The Initial Singularity}},
  Phys.\ Rev.\ Lett.\  {\bf 72}, 3305 (1994),
[\href{http://xxx.lanl.gov/abs/gr-qc/9312022}
{{\tt arXiv:gr-qc/9312022}}].

\bibitem{Starobinskii:1978} 
A.~A.~Starobinskii,
{\it{On a nonsingular isotropic cosmological model}},
Soviet Astronomy Letters, \textbf{4}, 82-84 (1978).


\bibitem{Mukhanov:1991zn}
  V.~F.~Mukhanov and R.~H.~Brandenberger,
 {\it{A Nonsingular universe}},
  Phys.\ Rev.\ Lett.\  {\bf 68}, 1969 (1992).

\bibitem{Novello:2008ra}
M.~Novello and S.~E.~P.~Bergliaffa,
{\it{Bouncing Cosmologies}},
Phys. Rept. \textbf{463}, 127-213 (2008)
         [\href{http://xxx.lanl.gov/abs/0802.1634}
{{\tt arXiv:0802.1634}}].







\bibitem{Veneziano:1991ek}
  G.~Veneziano,
 {\it{Scale Factor Duality For Classical And Quantum Strings}},
  Phys.\ Lett.\  B {\bf 265}, 287 (1991).



\bibitem{Khoury:2001wf}
  J.~Khoury, B.~A.~Ovrut, P.~J.~Steinhardt and N.~Turok,
 {\it{The ekpyrotic universe: Colliding branes and the origin of the hot
big bang}},
  Phys.\ Rev.\  D {\bf 64}, 123522
(2001),
[\href{http://xxx.lanl.gov/abs/hep-th/0103239}
{{\tt arXiv:hep-th/0103239}}].




 \bibitem{Khoury:2001bz}
  J.~Khoury, B.~A.~Ovrut, N.~Seiberg, P.~J.~Steinhardt and N.~Turok,
 {\it{From big crunch to big bang}},
  Phys.\ Rev.\  D {\bf 65}, 086007 (2002)
  [\href{http://xxx.lanl.gov/abs/hep-th/0108187}
{{\tt arXiv:hep-th/0108187}}].



\bibitem{Brustein:1997cv}
  R.~Brustein and R.~Madden,
 {\it{A model of graceful exit in string cosmology}},
  Phys.\ Rev.\  D {\bf 57}, 712 (1998),
[\href{http://xxx.lanl.gov/abs/hep-th/9708046}
{{\tt arXiv:hep-th/9708046}}].


\bibitem{Tirtho1}
  T.~Biswas, A.~Mazumdar and W.~Siegel,
 {\it{Bouncing universes in string-inspired gravity}},
  JCAP {\bf 0603}, 009 (2006),
[\href{http://xxx.lanl.gov/abs/hep-th/0508194}
{{\tt arXiv:hep-th/0508194}}].





\bibitem{Nojiri:2013ru}
  S.~Nojiri and E.~N.~Saridakis,
   {\it{Phantom without ghost}},
  Astrophys.\ Space Sci.\  {\bf 347}, 221 (2013)
  [\href{http://xxx.lanl.gov/abs/1301.2686}
{{\tt arXiv:1301.2686}}].

\bibitem{Starobinsky:1980te}
A.~A.~Starobinsky,
{\it{A New Type of Isotropic Cosmological Models Without Singularity}},
Phys. Lett. B \textbf{91} (1980), 99-102.

\bibitem{Bamba:2013fha}
  K.~Bamba, A.~N.~Makarenko, A.~N.~Myagky, S.~Nojiri and S.~D.~Odintsov,
  {\it{Bounce cosmology from $F(R)$ gravity and $F(R)$ bigravity}},
  JCAP {\bf 1401} (2014) 008
    [\href{http://xxx.lanl.gov/abs/1309.3748}
{{\tt arXiv:1309.3748}}].




\bibitem{Nojiri:2014zqa}
  S.~Nojiri and S.~D.~Odintsov,
   {\it{Mimetic $F(R)$ gravity: inflation, dark energy and bounce}},
  Mod.\ Phys.\ Lett.\ A {\bf 29}, no. 40, 1450211 (2014)
      [\href{http://xxx.lanl.gov/abs/1408.3561}
{{\tt arXiv:1408.3561}}].







 \bibitem{Shtanov:2002mb}
  Y.~Shtanov and V.~Sahni,
 {\it{Bouncing braneworlds}},
  Phys.\ Lett.\  B {\bf 557}, 1 (2003),
[\href{http://xxx.lanl.gov/abs/gr-qc/0208047}
{{\tt arXiv:gr-qc/0208047}}].



\bibitem{Saridakis:2007cf}
  E.~N.~Saridakis,
 {\it{Cyclic Universes from General Collisionless Braneworld Models}},
  Nucl.\ Phys.\ B {\bf 808}, 224 (2009),
[\href{http://xxx.lanl.gov/abs/0710.5269}
{{\tt arXiv:0710.5269}}].



\bibitem{Cai:2009in}
  Y.~F.~Cai and E.~N.~Saridakis,
 {\it{Non-singular cosmology in a model of non-relativistic gravity}},
  JCAP {\bf 0910}, 020 (2009),
[\href{http://xxx.lanl.gov/abs/0906.1789}
{{\tt arXiv:0906.1789}}].




\bibitem{Cai:2012ag}
  Y.~F.~Cai, C.~Gao and E.~N.~Saridakis,
 {\it{Bounce and cyclic cosmology in extended nonlinear massive gravity}},
  JCAP {\bf 1210}, 048 (2012)
  [\href{http://xxx.lanl.gov/abs/1207.3786}
{{\tt arXiv:1207.3786}}]




\bibitem{Cai:2010zma}
  Y.~-F.~Cai and E.~N.~Saridakis,
 {\it{Cyclic cosmology from Lagrange-multiplier modified gravity}},
  Class.\ Quant.\ Grav.\  {\bf 28}, 035010 (2011),
[\href{http://xxx.lanl.gov/abs/1007.3204}
{{\tt arXiv:1007.3204}}].

\bibitem{Ashtekar:2006wn}
  A.~Ashtekar, T.~Pawlowski and P.~Singh,
 {\it{Quantum Nature of the Big Bang: Improved dynamics}},
  Phys.\ Rev.\ D {\bf 74}, 084003 (2006)
  [\href{http://xxx.lanl.gov/abs/gr-qc/0607039}
{{\tt arXiv:gr-qc/0607039}}].








\bibitem{Bojowald:2001xe}
  M.~Bojowald,
 {\it{Absence of singularity in loop quantum cosmology}},
  Phys.\ Rev.\ Lett.\  {\bf 86}, 5227 (2001),
[\href{http://xxx.lanl.gov/abs/gr-qc/0102069}
{{\tt arXiv:gr-qc/0102069}}].

\bibitem{Ashtekar:2007em}
  A.~Ashtekar, A.~Corichi and P.~Singh,
 {\it{Robustness of key features of loop quantum cosmology}},
  Phys.\ Rev.\ D {\bf 77}, 024046 (2008)
  [\href{http://xxx.lanl.gov/abs/0710.3565}
{{\tt arXiv:0710.3565}}].







\bibitem{Minas:2019urp}
G.~Minas, E.~N.~Saridakis, P.~C.~Stavrinos and A.~Triantafyllopoulos,
{\it{Bounce cosmology in generalized modified gravities}},
Universe \textbf{5}, 74 (2019)
         [\href{http://xxx.lanl.gov/abs/1902.06558}
{{\tt arXiv:1902.06558}}].






\bibitem{Cai:2011tc}
  Y.~-F.~Cai, S.~-H.~Chen, J.~B.~Dent, S.~Dutta and E.~N.~Saridakis,
 {\it{Matter Bounce Cosmology with the f(T) Gravity}},
  Class.\ Quant.\ Grav.\  {\bf 28}, 215011 (2011),
[\href{http://xxx.lanl.gov/abs/1104.4349}
{{\tt arXiv:1104.4349}}].




\bibitem{Martin:2000xs}
J.~Martin and R.~H.~Brandenberger,
{\it{The TransPlanckian problem of inflationary cosmology}},
Phys. Rev. D \textbf{63}, 123501 (2001)
         [\href{http://xxx.lanl.gov/abs/hep-th/0005209}
{{\tt arXiv:hep-th/0005209}}].


\bibitem{Starobinsky:2001kn}
A.~A.~Starobinsky,
{\it{Robustness of the inflationary perturbation spectrum to transPlanckian physics}},
JETP Lett. \textbf{73},371-374 (2001) 
[\href{https://arxiv.org/abs/astro-ph/0104043}
{{\tt arXiv:astro-ph/0104043}}].

\bibitem{Starobinsky:2002rp}
A.~A.~Starobinsky and I.~I.~Tkachev,
{\it{Trans-Planckian particle creation in cosmology and ultra-high energy cosmic rays}},
JETP Lett. \textbf{76}, 235-239 (2002) 
[\href{https://arxiv.org/abs/astro-ph/0207572}
{{\tt arXiv:astro-ph/0207572}}].

\bibitem{Brandenberger:2012aj}
R.~H.~Brandenberger and J.~Martin,
{\it{Trans-Planckian Issues for Inflationary Cosmology}},
Class. Quant. Grav. \textbf{30}, 113001 (2013)
         [\href{http://xxx.lanl.gov/abs/1211.6753}
{{\tt arXiv:1211.6753}}].




\bibitem{Wands:1998yp}
D.~Wands,
{\it{Duality invariance of cosmological perturbation spectra}},
Phys. Rev. D \textbf{60}, 023507 (1999)
         [\href{http://xxx.lanl.gov/abs/gr-qc/9809062}
{{\tt arXiv:gr-qc/9809062}}].




\bibitem{Finelli:2001sr}
F.~Finelli and R.~Brandenberger,
{\it{On the generation of a scale invariant spectrum of adiabatic fluctuations
in
cosmological models with a contracting phase}},
Phys. Rev. D \textbf{65}, 103522 (2002)
         [\href{http://xxx.lanl.gov/abs/hep-th/0112249}
{{\tt arXiv:hep-th/0112249}}].




\bibitem{Biswas:2015kha}
T.~Biswas, R.~Mayes and C.~Lattyak,
{\it{Perturbations in bouncing and cyclic models}},
Phys. Rev. D \textbf{93}, no.6, 063505 (2016)
         [\href{http://xxx.lanl.gov/abs/1502.05875}
{{\tt arXiv:1502.05875}}].


\bibitem{Starobinsky:1979ty}
A.~A.~Starobinsky,
{\it{Spectrum of relict gravitational radiation and the early state of the universe}},
Pisma Zh.Eksp.Teor.Fiz. 30 (1979) 719-723 
[JETP Lett. \textbf{30}, 682-685 (1979)].


\bibitem{Bamba:2016gbu}
K.~Bamba, G.~G.~L.~Nashed, W.~El Hanafy and S.~K.~Ibraheem,
{\it{Bounce inflation in $f(T)$ Cosmology: A unified inflaton-quintessence
field}},
Phys. Rev. D \textbf{94}, no.8, 083513 (2016)
         [\href{http://xxx.lanl.gov/abs/1604.07604}
{{\tt arXiv:1604.07604}}].





\bibitem{ElHanafy:2017xsm}
W.~El Hanafy and G.~G.~L.~Nashed,
{\it{Generic phase portrait analysis of finite-time singularities and
generalized Teleparallel gravity}},
Chin. Phys. C \textbf{41}, no.12, 125103 (2017)
         [\href{http://xxx.lanl.gov/abs/1702.05786}
{{\tt arXiv:1702.05786}}].



\bibitem{ElHanafy:2017sih}
W.~El Hanafy and G.~G.~L.~Nashed,
{\it{Lorenz Gauge Fixing of $f(T)$ Teleparallel Cosmology}},
Int. J. Mod. Phys. D \textbf{26}, no.14, 1750154 (2017)
         [\href{http://xxx.lanl.gov/abs/1707.01802}
{{\tt arXiv:1707.01802}}].




\bibitem{Awad:2017yod}
A.~Awad, W.~El Hanafy, G.~G.~L.~Nashed and E.~N.~Saridakis,
{\it{Phase Portraits of general $f(T)$ Cosmology}},
JCAP \textbf{02}, 052 (2018)
         [\href{http://xxx.lanl.gov/abs/1710.10194}
{{\tt arXiv:1710.10194}}].

\bibitem{book:Steven}
Steven H. Strogatz,
{\it Nonlinear Dynamics And Chaos: With Applications To Physics,
Biology, Chemistry And Engineering},
Westview Press, Colorado (2015).




\bibitem{Awad:2013tha}
A.~Awad,
{\it{Fixed points and FLRW cosmologies: Flat case}},
Phys. Rev. D \textbf{87}, no.10, 103001 (2013)
[erratum: Phys. Rev. D \textbf{87}, no.10, 109902 (2013)]
         [\href{http://xxx.lanl.gov/abs/1303.2014}
{{\tt arXiv:1303.2014}}].





\bibitem{Capozziello:2015rda}
S.~Capozziello, O.~Luongo and E.~N.~Saridakis,
{\it{Transition redshift in $f(T)$ cosmology and observational constraints}},
Phys. Rev. D \textbf{91}, no.12, 124037 (2015)
         [\href{http://xxx.lanl.gov/abs/1503.02832}
{{\tt arXiv:1503.02832}}].


\bibitem{Ellis:2002we}
G.~F.~R.~Ellis and R.~Maartens,
{\it{The emergent universe: Inflationary cosmology with no singularity}},
Class. Quant. Grav. \textbf{21}, 223-232 (2004)
         [\href{http://xxx.lanl.gov/abs/gr-qc/0211082}
{{\tt arXiv:gr-qc/0211082}}].


\bibitem{Mulryne:2005ef}
D.~J.~Mulryne, R.~Tavakol, J.~E.~Lidsey and G.~F.~R.~Ellis,
{\it{An Emergent Universe from a loop}},
Phys. Rev. D \textbf{71}, 123512 (2005)
         [\href{http://xxx.lanl.gov/abs/astro-ph/0502589}
{{\tt arXiv:astro-ph/0502589}}].





\bibitem{Bunch:1978yq}
T.~S.~Bunch and P.~C.~W.~Davies,
{\it{Quantum Field Theory in de Sitter Space: Renormalization by Point
Splitting}},
Proc. Roy. Soc. Lond. A \textbf{360}, 117-134 (1978).

\bibitem{Keresztes:2012zn}
Z.~Keresztes, L.~A.~Gergely and A.~Y.~Kamenshchik,
{\it{The paradox of soft singularity crossing and its resolution by
distributional
cosmological quantitities}},
Phys. Rev. D \textbf{86}, 063522 (2012)
         [\href{http://xxx.lanl.gov/abs/1204.1199}
{{\tt arXiv:1204.1199}}].







\bibitem{Cai:2011bs}
Y.~F.~Cai and E.~N.~Saridakis,
{\it{Non-singular Cyclic Cosmology without Phantom Menace}},
J. Cosmol. \textbf{17}, 7238-7254 (2011)
         [\href{http://xxx.lanl.gov/abs/1108.6052}
{{\tt arXiv:1108.6052}}].




\bibitem{Weinberg:1972kfs}
S.~Weinberg,
{\it{Gravitation and Cosmology: Principles and Applications of the General
Theory
of Relativity}}, John Wiley \& Sons, New York (1972).
%

\bibitem{Farooq:2016zwm}
O.~Farooq, F.~R.~Madiyar, S.~Crandall and B.~Ratra,
{\it{Hubble Parameter Measurement Constraints on the Redshift of the
Deceleration\textendash{}acceleration Transition, Dynamical Dark Energy, and
Space Curvature}},
Astrophys. J. \textbf{835}, no.1, 26 (2017)
         [\href{http://xxx.lanl.gov/abs/1607.03537}
{{\tt arXiv:1607.03537}}].



\bibitem{Frampton:2011aa}
P.~H.~Frampton, K.~J.~Ludwick and R.~J.~Scherrer,
{\it{Pseudo-Rip: Cosmological models intermediate between the cosmological
constant and the little rip}},
Phys. Rev. D \textbf{85}, 083001 (2012)
         [\href{http://xxx.lanl.gov/abs/1112.2964}
{{\tt arXiv:1112.2964}}].




\bibitem{delaCruz-Dombriz:2014zaa}
\'A.~de la Cruz-Dombriz, P.~K.~S.~Dunsby and D.~Saez-Gomez,
{\it{Junction conditions in extended Teleparallel gravities}},
JCAP \textbf{12} (2014), 048
doi:10.1088/1475-7516/2014/12/048
[\href{https://arxiv.org/abs/1406.2334}
{{\tt arXiv:1406.2334}}].

\bibitem{Awad:2017sau}
A.~Awad and G.~Nashed,
{\it{Generalized teleparallel cosmology and initial singularity crossing}},
JCAP \textbf{02} (2017), 046
doi:10.1088/1475-7516/2017/02/046
 [\href{https://arxiv.org/abs/1701.06899}
{{\tt arXiv:1701.06899}}].


\bibitem{Weitzenb23}
  Weitzenb\"{o}ck R.,
  \emph{Invarianten Theorie},
  Nordhoff, Groningen (1923).





\bibitem{Chavanis:2012pd}
P.~H.~Chavanis,
 {\it{ Models of universe with a polytropic equation of state: I. The early
universe,}}
Eur. Phys. J. Plus \textbf{129}, 38 (2014)
         [\href{http://xxx.lanl.gov/abs/1208.0797}
{{\tt arXiv:1208.0797}}].

\bibitem{Nojiri:2005sx}
S.~Nojiri, S.~D.~Odintsov and S.~Tsujikawa,
{\it{Properties of singularities in (phantom) dark energy universe}},
Phys. Rev. D \textbf{71}, 063004 (2005)
         [\href{http://xxx.lanl.gov/abs/hep-th/0501025}
{{\tt arXiv:hep-th/0501025}}].



\bibitem{Frampton:2011rh}
P.~H.~Frampton, K.~J.~Ludwick, S.~Nojiri, S.~D.~Odintsov and R.~J.~Scherrer,
{\it{Models for Little Rip Dark Energy}},
Phys. Lett. B \textbf{708}, 204-211 (2012)
         [\href{http://xxx.lanl.gov/abs/1108.0067}
{{\tt arXiv:1108.0067}}].




\bibitem{Saridakis:2021qxb}
E.~N.~Saridakis,
{\it{Do we need soft cosmology?}},
         [\href{http://xxx.lanl.gov/abs/2105.08646}
{{\tt arXiv:2105.08646}}].

 


\end{thebibliography}
\end{document}